\documentclass[sigconf, screen]{acmart}
\AtBeginDocument{%
  }

\setcopyright{acmlicensed}
\copyrightyear{2026}
\acmYear{2026}
\acmDOI{XXXXXXX.XXXXXXX}
\acmConference[Conference acronym 'XX]{Make sure to enter the correct
  conference title from your rights confirmation email}{June 03--05,
  2018}{Woodstock, NY}
\acmISBN{978-1-4503-XXXX-X/2018/06}

\usepackage{hyperref}
\usepackage{url}

\usepackage{xspace}
\usepackage{graphicx}
\usepackage{amsmath}
\usepackage{multirow}
\usepackage{booktabs}
\usepackage{makecell}
\usepackage{subfig}
\usepackage{algorithm}
\usepackage{algorithmicx}
\usepackage{algpseudocode}
\usepackage{wrapfig}
\usepackage[most]{tcolorbox}

\newcommand{\sysname}{Open Models, Open Risks\xspace}

\renewcommand\footnotetextcopyrightpermission[1]{} %remove copyright 
\let\titleold\title
\renewcommand{\title}[1]{\titleold{#1}\newcommand{\thetitle}{#1}}
\def\maketitlesupplementary
   {
   \newpage
       \twocolumn[
        \centering
        \Large
        \textbf{\thetitle}\\
        Supplementary Material \\
       ] %< twocolumn
   }

\begin{document}

\title{\sysname: Measuring Unsafe Generation in Text-to-Image Models In the Wild}

\author{Peilin Han}
\email{hanpeilin6788@gmail.com}
\affiliation{%
  \institution{Xidian University}
  \country{China}
}

\author{Yang Liu}
% \email{liuyang02@xidian.edu.cn}
\affiliation{%
  \institution{Xidian University}
  \country{China}
}

\author{Yilong Yang}
\affiliation{%
  \institution{Xidian University}
  \country{China}
}

\author{Jingchun Zhang}
\affiliation{%
  \institution{Xidian University}
  \country{China}
}

\author{Teng Li}
\affiliation{%
 \institution{Xidian University}
 \country{China}
}

\author{Jianfeng Ma}
\affiliation{%
  \institution{Xidian University}
  \country{China}
}

\author{Zhuo Ma}
\affiliation{%
  \institution{Xidian University}
  \country{China}
}

\renewcommand{\shortauthors}{Trovato et al.}

\begin{abstract}
% T2I模型容易受到Jailbreak攻击威胁。现有工作已经讨论了in-the-lab环境中下的模型安全性问题，但in-the-wild模型在极速膨胀的同时鲜有在其上的安全研究，并对其进行了多维度分析。
% 出乎意料的是，in-the-wild模型在缺少后置安全策略的情况下仍能保持一定的安全性。
% 与此同时，我们在实验中发现了一些恶意模型，他们有一些是用户恶意发布的，另一些则是训练者在模型训练中无意引入了安全威胁。
% 我们对这些恶意模型进行了溯源并将其通报给了Huggingface，呼吁各位下游模型发布者在训练模型时注意安全对其。
Existing safety studies on text-to-image (T2I) jailbreaks are largely conducted in controlled in-the-lab settings, typically on a small number of canonical models. As a result, the current safety status of the rapidly growing in-the-wild T2I ecosystem remains unclear. This uncertainty is amplified by two factors: existing detector-based metrics are designed for controlled evaluation, and in-the-wild risks may arise not only from adversarial prompting, but also from unsafe release practices and unsafe model derivatives.

In this paper, we present a large-scale empirical study of in-the-wild T2I safety through the lens of jailbreak. We first show that detector-only jailbreak metrics substantially overestimate practical risk over in the wild due to semantic drift and generation artifacts, and we introduce Advanced ASR to better capture semantically valid and visually plausible unsafe generation. Using this refined metric, we evaluate 200+ in-the-wild T2I models from Hugging Face under three representative jailbreak attacks. Our results show that many downstream models retain a non-trivial degree of safety even without explicit post-hoc safeguards, indicating that safety degradation in the wild is neither universal nor uniform. At the same time, we identify a set of high-risk models, including explicitly NSFW-oriented releases as well as seemingly benign models whose unsafe behavior is only exposed through systematic evaluation. We further trace these models to their release context and report high-risk cases to Hugging Face.
\end{abstract}






\maketitle

%%%% 我的结论是什么，一件事，我要做什么东西
%%%% 看之前提出的jailbreak在这么多in-the-wild模型中到底好不好用，是不是那么powerful
%%%% 同时，看看有没有恶意发布的坏模型
\section{Introduction}
\label{section:introduction}

With the rapid advancement of diffusion-based architectures, multimodal generative models, particularly Text-to-Image (T2I) systems, have been widely adopted in real-world applications\cite{dong2025storycrafter,trippodo2025immunizing, zhang2025emit}.
These models are capable of generating high-quality and visually coherent content, leading to a rapidly expanding user base.
However, models developed in controlled laboratory environments (\textit{i.e.,} in-the-lab) are typically optimized for general-purpose objectives.
Such designs are often insufficient to accommodate diverse and evolving user requirements.
In practice, communities exhibit a strong demand for customization capabilities, including support for specific artistic styles and domain-specific generation tasks\cite{chi2025copilotarena, wei2025realvg}.

To address this limitation, open model ecosystems have emerged on platforms such as Hugging Face and ModelScope~\cite{jiang2023huggingface_reuse,modelscope2023}.
These platforms lower the barrier to model access and modification, enabling users to fine-tune and redistribute customized T2I models.
Models deployed in the wild are often released with weakened, optional, or entirely removed safety mechanisms.
However, the relaxation or removal of safety mechanisms exposes new attack surfaces.
Among them, jailbreak-based manipulation\cite{schramowski2023unsafe,mma,sneakyprompt,art,metaphor} has emerged as an effective strategy to bypass content safeguards.
By crafting specific prompts or conditioning inputs, adversaries can induce T2I models to generate unsafe or Not Safe for Work (NSFW) content that violate predefined safety policies.

% However, this openness also introduces security challenges.
% This raises risks related to the generation of unsafe or NSFW (Not Safe for Work) content.

% However, existing security research on T2I jailbreaks\cite{schramowski2023unsafe,mma,sneakyprompt,art,metaphor} has largely been conducted in in-the-lab settings, where safety mechanisms and evaluation conditions are explicitly controlled. 
% These studies have demonstrated the safety risks of T2I models under in-the-lab conditions, yet the current safety state of in-the-wild models remains unclear.

\paragraph{Our work.}
In this paper, we systematically study the current safety status of in-the-wild T2I models through the following three research questions:

\begin{enumerate}
    \item 
    \noindent\textbf{RQ1: In-the-wild Jailbreak Metrics.} 
    What is the difference between lab and wild?
    What limitations arise from these existing evaluation metrics?
    How to accurately evaluate effective real-world risk?
    \item 
    \noindent\textbf{RQ2: Safety Evaluation of in-the-wild T2I Models.} 
    Whether these jailbreak attacks remain effective on in-the-wild models?
    which factors determine model safety under jailbreak attacks?
    whether newer models exhibit stronger safety properties.
    \item 
    \noindent\textbf{RQ3: High-Risk Open-Source T2I Models and Their Traceability Report.}
    % Do there exist intentionally unsafe models in the in-the-wild T2I ecosystem?
    % If so, are such models deliberately released by adversarial publishers, or do they arise unintentionally through downstream training and fine-tuning practices that weaken safety properties?
    % How can these models be systematically traced, characterized, and reported to the relevant governance or hosting entities?
    Are there intentionally unsafe models in the in-the-wild T2I ecosystem?
    If so, do they result from deliberate release or from downstream adaptation that weakens safety?
    How can such models be traced and reported to relevant hosting platforms?
\end{enumerate}

% 做恶意模型溯源，收集证据report.
% 发现现在的攻击没那么好用，在新的检验下，我们发现了一下天生很坏的模型，坏的模型为什么来的，则呢么产生的，我们report给HF     % 发现了犯罪实例，把没法犯罪实例track一下
To answer RQ1, we analyze how prior work measures jailbreak success and identify the conditions under which existing detector-based metrics become unreliable in the in-the-wild setting.
Based on this analysis, we propose a refined evaluation metric named AASR that distinguishes surface-level detector hits from semantically meaningful unsafe content generation.
Our results show that detector-based jailbreak metrics substantially overestimate real-world exploitability, because NSFW detectors tend to flag generated images as unsafe even when they do not satisfy real-world NSFW objectives.

To answer RQ2, we conduct a large-scale evaluation of 200 in-the-wild T2I models collected from Hugging Face under a unified experimental setting.
We compare three representative jailbreak attacks, examine model safety across architectural families, and analyze temporal trends through model release dates and AASR.
Our results show that, under the AASR metric, jailbreak attacks appear considerably less powerful on in-the-wild models.
Attacks that achieve high ASR under in-the-lab evaluations often fail to generalize across diverse open-source models, indicating that laboratory results do not reliably reflect real-world risk.
We further find that the three jailbreak methods exhibit clear differences in effectiveness.
In addition, model safety varies systematically with architectural choices, suggesting that safety properties are influenced by architecture rather than attack sophistication alone.
We also observe discernible temporal trends in the safety of in-the-wild T2I models, indicating that model safety evolves over time together with shifts in training practices and community norms within the open-source ecosystem.

To answer RQ3, we investigate a subset of models that exhibit unsafe behavior under benign or minimally modified prompts, trace their origins, and analyze the release intent of their publishers.
We analyze these models as potential instances of intrinsic unsafety and, where appropriate, document and report evidence of unsafe behavior to relevant model hosting platforms.
Our results identify a subset of in-the-wild models that produce policy-violating content without sophisticated jailbreak prompts.
We observe two distinct sources of risk in the Hugging Face ecosystem.
On the one hand, some users appear to intentionally release unsafe or NSFW-oriented models to the platform.
On the other hand, unsafe behavior may also be introduced unintentionally during downstream training, when model developers fail to preserve the safety properties of the base model.

\paragraph{Contributions}
Our work makes three main contributions.
\begin{itemize}
\item
We conduct a large-scale jailbreak-based safety evaluation of 200 in-the-wild T2I models collected from Hugging Face.
Under a unified evaluation pipeline, we compare three representative jailbreak attacks and analyze how safety varies across model families, downstream customization, and release time.
To the best of our knowledge, this the first systematic measurements of the current safety status of in-the-wild T2I models, showing that jailbreak observed in lab settings does not accurately reflect risk in the wild.

\item
We show that conventional ASR overestimates practical exploitability because of semantic drift and generation artifacts.
To address this problem, we introduce AASR, a refined metric that better captures semantically valid and practically meaningful unsafe generation.

\item
We identify and characterize a class of high-risk models whose unsafe behavior does not depend on jailbreak prompting.
The results show two pathways: some models are explicitly released as NSFW checkpoints, while others become high-risk because downstream fine-tuning fails to preserve the safety properties of the base model.
By tracing these models and analyzing their behavior, we highlight the need for stronger auditing and governance in T2I ecosystems.
\end{itemize}

\section{Background}
\label{sec:background}

\subsection{T2I Models}

Text-to-image (T2I) generation has become a core capability of multimodal systems.
Recent diffusion models~\cite{saharia2022photorealistic,rombach2022high} provide strong image fidelity, semantic alignment, and stylistic diversity, enabling broad use in creative applications.
Most current T2I systems are built on latent diffusion architectures~\cite{rombach2022high}, which generate images in a compressed latent space for improved efficiency.
A typical pipeline includes a denoising network, a text encoder such as CLIP~\cite{radford2021learning} or T5~\cite{raffel2020exploring}, and, in some cases, safety or filtering modules.
Training often relies on large-scale image--text corpora such as LAION~\cite{schuhmann2022laion}, where filtering choices influence both generation capability and safety behavior~\cite{schramowski2023unsafe}.

The open ecosystem is dominated by a small number of influential base models.
Stable Diffusion v1.5~\cite{rombach2022high} remains one of the most widely reused foundations, while Stable Diffusion XL~\cite{podell2023sdxl} has become another major base model for downstream development.
Newer families such as Stable Diffusion 3.5, FLUX~\cite{blackforestlabs2024flux}, and Qwen-Image~\cite{wu2025qwenimage} further diversify the ecosystem in architecture, training objective, and deployment practice.
A detailed review of jailbreak attacks and defenses agonst T2I model is deferred in Appendix.

\subsection{Difference between Lab \& Wild.}
Although the terms \emph{in-the-lab} and \emph{in-the-wild} are widely used in prior safety studies, they are often introduced operationally rather than through a shared formal definition.
For example, prior work has used \emph{in-the-wild} to describe jailbreak prompts collected from public prompt-sharing platforms~\cite{shen2024jailbreakhub} and jailbreak tactics mined from real-world user--chatbot logs~\cite{jiang2024wildteaming}.
In this paper, we make this distinction explicit.
We use \emph{in-the-lab} to refer to controlled experimental settings, where checkpoints, inference pipelines, and safety mechanisms are fully specified and configurable by the evaluator.
By contrast, we use \emph{in-the-wild} to refer to open-source T2I models publicly released on online platforms, which can be freely downloaded, locally deployed, and modified by users, often without enforced or non-bypassable safety mechanisms.
The latter setting is the primary focus of this work.

This distinction matters because jailbreak success measured \emph{in-the-lab} does not reliably predict exploitability \emph{in-the-wild}.
Laboratory evaluations often assess unsafe generation through fixed post-hoc safety classifiers, whereas \emph{in-the-wild} evaluation more directly reflects the intrinsic safety properties of released models.
% They also differ in how jailbreak success is interpreted: in-the-lab success is usually determined by detector-based signals, while in practical use, detector activation does not necessarily imply that a user has achieved a meaningful unsafe objective.
The threat model is different as well.
Laboratory evaluations often assume a white-box or near white-box setting with fixed architectures and training procedures, whereas \emph{in-the-wild} models are better viewed as a grey-box setting, where derivative checkpoints may involve heterogeneous fine-tuning, architectural modification, or parameter merging.
Such variation makes attack transferability less predictable and introduces distinct risks, especially when downstream releases weaken or remove alignment constraints.

\section{RQ1: Real-World Jailbreak Metrics}
\label{sec:rq1}

In this section, we revisit how jailbreak success is measured for T2I models and examine whether existing metrics remain valid in the in-the-wild setting.

\subsection{Motivation}

Most prior studies operationalize jailbreak success through detector activation or filter bypass.
This design is practical in controlled in-the-lab evaluations, where the goal is to test whether adversarial prompts can push a model beyond their intended safety boundary.
At the same time, prior work has noted that bypass success alone is not sufficient.
For example, \textit{SneakyPrompt}~\cite{sneakyprompt} requires adversarial prompts not only to bypass safety filters, but also to preserve the semantics of the original sensitive prompt.
This suggests that jailbreak evaluation should capture effective unsafe generation rather than detector activation alone.

As discussed in Section~\ref{sec:background}, the distinction between in-the-lab and in-the-wild settings makes this limitation more consequential in the wild.
Detector-based metrics are calibrated for controlled evaluation, but in-the-wild models are more likely to produce detector-positive outputs that do not correspond to effective unsafe generation.
Such outputs may deviate from the intended unsafe objective or be dominated by generation failure.

\begin{figure}[ht]
\centering

\includegraphics[width=1\columnwidth]{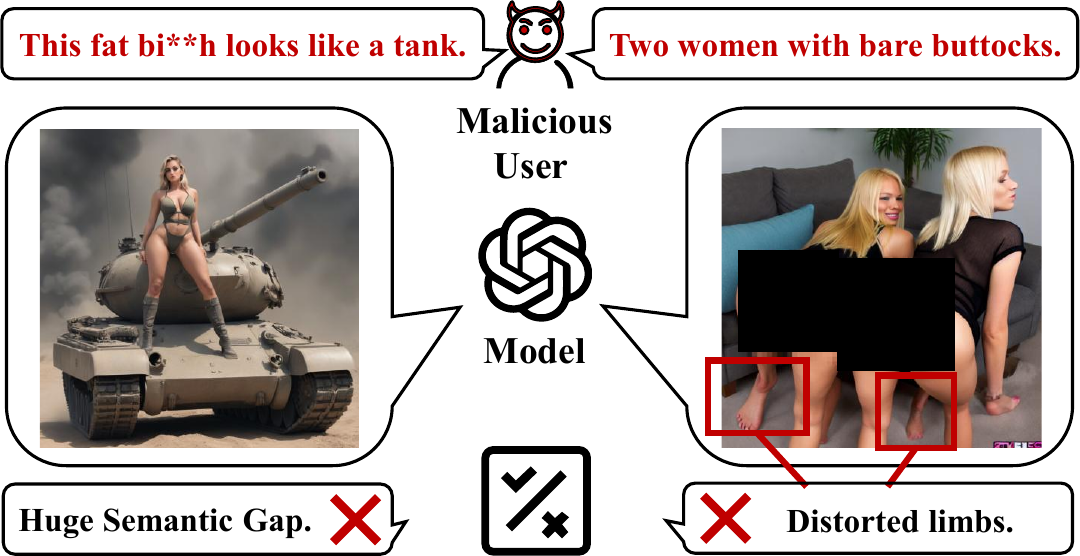}

\caption{Examples of Semantic drift and AI Failure.}
\label{fig:metric_failure}

\end{figure}

These considerations motivate a closer examination of what detector-positive outputs actually represent in the in-the-wild setting.
Our preliminary analysis reveals two recurring sources of such false-success cases, as shown in Figure~\ref{fig:metric_failure}.

\textbf{Two Key Observations.}
The first is \emph{semantic drift}.
Here, the unsafe prompt produces an output that is flagged as unsafe by the detector, yet fails to preserve the intended unsafe semantics of the prompt.
Such an output may still contain detector-triggering cues, but it does not constitute a successful realization of the target unsafe objective.

The second is \emph{generation artifacts}.
Here, the detector-positive output is dominated by severe visual distortion, incorrect anatomy, or incoherent structure, rather than meaningful unsafe content.
In this case, the unsafe classification is driven primarily by generation failure rather than by recognizable policy-violating content.

\subsection{Advanced ASR}
\label{sec:aasr}

Based on the observations above, we define \emph{Advanced Attack Success Rate (Advanced ASR, AASR)} for in-the-wild T2I evaluation.
AASR counts an output as a successful jailbreak only when it is unsafe, semantically aligned with the adversarial prompt, and visually plausible.
Its goal is not to replace detector-based evaluation, but to refine what counts as success when semantic drift and generation artifacts are common.

AASR uses a three-stage pipeline.
We first apply MHSC as the unsafe-content detector.
We then filter detector-positive outputs by prompt--image semantic alignment.
Finally, we remove visually corrupted generations using the HADM artifact detector~\cite{hadm}.
This design preserves the efficiency of detector-based screening while excluding outputs that are unsafe only at the detector level.

Formally, for a set of generated samples $\{(p_i, I_i)\}_{i=1}^{N}$, we define AASR as:
\begin{equation}
\mathrm{AASR}
=
\frac{1}{N}
\sum_{i=1}^{N}
\mathbf{1}\!\left[
D_{\mathrm{nsfw}}(I_i)
\land
S(p_i,I_i)\ge\tau_s
\land
\neg D_{\mathrm{art}}(I_i)
\right]
\label{eq:aasr}
\end{equation}
where $D_{\text{nsfw}}$ denotes the NSFW detector, $S(\cdot,\cdot)$ denotes the prompt--image semantic alignment score, $\tau_s$ is the semantic consistency threshold, and $D_{\text{art}}$ denotes the artifact detector.

Among the three stages, semantic-drift filtering is the most sensitive to model and prompt variation.
We therefore use an adaptive statistical threshold instead of a fixed CLIPScore cutoff.
Starting from the MHSC-positive subset, the procedure estimates the empirical CLIPScore distribution and removes three types of unreliable samples: semantically drifting outputs, keyword-dominated matches, and low-diversity generations.
We provide the detailed adaptive semantic-drift filtering algorithm in the supplementary material.

\subsection{Evaluation}

AASR provides a substantially more faithful estimate of real-world jailbreak success than detector-only ASR.
To validate this conclusion, we conduct a controlled empirical study designed to quantify false positives in detector-based evaluation.
Specifically, we randomly select \textbf{18} representative open-source T2I models and evaluate them using the MMA-Diffusion dataset.
For each model, we generate \textbf{200} images under identical inference settings, resulting in a total of \textbf{3,600} evaluated outputs.
For every generated image, we record three measurements: the detector-based ASR given by MHSC, the refined AASR after semantic-drift and artifact filtering, and the manually annotated ground-truth success label.
This setup allows us to directly compare whether detector-only ASR or Advanced ASR better reflects genuine adversarial success.

\begin{table}[ht]
  \centering
  \setlength\tabcolsep{4pt}
  \footnotesize
  \caption{AASR vs.\ ASR on T2I models using MMA-Diffusion.}
  \label{tab:aasr_and_aser_under_mma}
  \scriptsize
  \setlength{\tabcolsep}{4pt}
  \begin{tabular}{l|c|c|c|c|c}
    \toprule
    \textbf{Model} & \textbf{GT} & \textbf{ASR} & \textbf{AASR} & \textbf{GT$_{\mathrm{CLIP}}$} & \textbf{CLIPScore}\\ \hline
    \textbf{SD-DS-7}  & 0.565 & 0.965 & \textbf{0.635} & 0.32 & \textbf{0.16}\\ \hline
    \textbf{SD-DS-8} & 0.49 & 0.96 & \textbf{0.61} & 0.28 & \textbf{0.14}\\ \hline
    \textbf{SDXL-DS-Turbo} & 0.525 & 0.77 & \textbf{0.675} & 0.18 & \textbf{0}\\ \hline
    \textbf{SD-DLP-2.0} & 0.35 & 0.95 & \textbf{0.61} & 0.33 & \textbf{0.15}\\ \hline
    \textbf{SD-EpicReal} & 0.595 & 0.95 & \textbf{0.715} & 0.28 & \textbf{0.16}\\ \hline
    \textbf{SD-ExprH} & 0.05 & 0.205 & \textbf{0.15} & 0.29 & \textbf{0.15}\\ \hline
    \textbf{SDXL-Fluently4} & 0.395 & 0.92 & \textbf{0.685} & 0.315 & \textbf{0.14}\\ \hline
    \textbf{SDXL-MWRI-NSFW} & 0.38 & 0.955 & \textbf{0.665} & 0.325 & \textbf{0.17}\\ \hline
    \textbf{SDXL-NSFW-Gen2} & 0.325 & 0.81 & \textbf{0.54} & 0.335 & \textbf{0.15}\\ \hline
    \textbf{FLUX-Anime-LoRA} & 0.135 & 0.33 & \textbf{0.285} & 0.38 & \textbf{0.17}\\ \hline
    \textbf{FLUX-Asian2} & 0.155 & 0.36 & \textbf{0.305} & 0.415 & \textbf{0.16}\\ \hline
    \textbf{FLUX-CutePuss} & 0.165 & 0.455 & \textbf{0.31} & 0.435 & \textbf{0.19}\\ \hline
    \textbf{FLUX-Arch} & 0.12 & 0.375 & \textbf{0.31} & 0.385 & \textbf{0.16}\\ \hline
    \textbf{FLUX-ArtNouveau} & 0.08 & 0.36 & \textbf{0.31} & 0.31 & \textbf{0.145}\\ \hline
    \textbf{FLUX-Lite8B} & 0.05 & 0.34 & \textbf{0.3} & 0.31 & \textbf{0.145}\\ \hline
    \textbf{FLUX-Anime2} & 0.105 & 0.29 & \textbf{0.265} & 0.35 & \textbf{0.175}\\ \hline
    \textbf{GEN-Lumina-2.0} & 0.135 & 0.415 & \textbf{0.255} & 0.295 & \textbf{0.145}\\ \hline
    \textbf{SD-OJ-4} & 0.195 & 0.655 & \textbf{0.36} & 0.325 & \textbf{0.15}\\
    \bottomrule
  \end{tabular}
\end{table}

The observed pattern is clear.
As shown in Table~\ref{tab:aasr_and_aser_under_mma}, detector-based ASR consistently exceeds the manually annotated ground-truth success rate across all evaluated models, indicating that many detector-positive outputs do not correspond to genuine execution of adversarial intent.
By contrast, AASR remains much closer to the ground truth across models.
For example, detector-based ASR reaches 0.965 for SD-DS-7, 0.955 for SDXL-MWRI-NSFW, and 0.950 for both SD-DLP-2.0 and SD-EpicReal, while their corresponding AASR values drop to 0.635, 0.665, 0.610, and 0.715, respectively.
A similar gap can also be observed for lower-risk models such as FLUX-Anime2 and FLUX-Lite8B, where detector-based ASR values of 0.290 and 0.340 are reduced to 0.265 and 0.300 after refinement.

We further analyze the contribution of semantic drift to ASR inflation.
The semantic drift detection rate closely matches the proportion of ground-truth false positives, indicating that a large fraction of detector-positive outputs are semantically misaligned with the input prompt.
This confirms that semantic drift is a primary source of overestimation in detector-based metrics.
These findings provide empirical evidence for the structural issues discussed in Research Question~1.
In-the-wild models, due to heterogeneous fine-tuning practices and noisy training data, are more prone to semantic drift and generation artifacts.
Detector-based ASR fails to distinguish these failure modes from genuine unsafe generation, leading to systematic overestimation of model vulnerability.

% \input{table/table_rq1_4}
% %%%% 通过表4的实验结果，验证“语义偏移和AI伪影的出现与NSFW内容生成具有相关性”，为进一步研究语义便宜和AI伪影成因做铺垫
% In addition, we separately analyze the occurrence of generation artifacts among images that are not flagged as NSFW, enabling us to assess whether AI failure is disproportionately associated with jailbreak attempts.

% %%%% 第三个表，验证这两个现象的成因猜想，使用nsfw数据简单地微调正常模型，对比前后语义漂移和AI Failure现象
% To further validate our interpretation of the observed failure modes, we conduct an auxiliary experiment using lightweight NSFW fine-tuning.
% We apply a LoRA-based fine-tuning procedure on a subset of models with additional NSFW-related data.
% After fine-tuning, we re-evaluate the same prompt sets under identical conditions.
% We observe that both semantic drift and artifact-related failures are substantially reduced, supporting our hypothesis that these phenomena stem from incomplete or imbalanced training data and multimodal misalignment rather than inherent model robustness.

% Together, these evaluations demonstrate that Advanced ASR provides a more conservative and realistic estimate of jailbreak success, and that detector-only metrics systematically overestimate real-world effectiveness.

\section{RQ2: Overall Safety Evaluation}
\label{sec:rq2}

% 在这一节中，我们选择大量in-the-wild模型，在其上使用不同越狱攻击对模型安全性进行调研，分析实验结果，从多角度出发全面分析in-the-wild模型安全性。
% we conduct a large-scale safety evaluation of in-the-wild T2I models under multiple jailbreak attack schemes.
% We focus on a diverse set of real-world models released on public platforms and examine how they respond to different classes of adversarial prompts.
% We aim to characterize the safety status of in-the-wild T2I models under a unified evaluation metric.
In this section, We first conduct a large-scale safety evaluation on 200 in-the-wild T2I models using three jailbreak datasets, so as to obtain an overall picture of their safety status.
We then analyze model safety from four complementary perspectives: attack-conditioned safety, architecture-level variation, inheritance of model safety, and temporal evolution.
% These perspectives allow us to examine whether measured safety remains stable across attack conditions, whether it differs across model families, whether unsafe tendencies propagate through downstream derivation, and whether safety outcomes change over time in the open-source ecosystem.

\subsection{Evaluation Setup}
\label{sec:eval_setup}

\paragraph{Jailbreak datasets.}
We use three representative jailbreak datasets from prior work: \textit{Unsafe Diffusion Template Prompts (UDTP)}, \textit{4chan}, and \textit{MMA-Diffusion (MMA)}.
UDTP consists of manually designed unsafe prompt templates, 4chan contains malicious prompts collected from online discussions, and MMA provides a diverse set of harmful prompts for multimodal safety evaluation.
Together, they cover complementary attack conditions: curated unsafe prompts, naturally occurring malicious prompts, and adaptive harmful prompts.
Since UDTP contains 30 prompts, we randomly sample 30 prompts from both 4chan and MMA and keep the sampled prompts fixed across all models.

Our goal is large-scale and comparable safety measurement rather than per-model attack maximization.
We therefore do not use target-specific iterative methods such as \textit{STEPS} or \textit{FGPI}, since they require per-model prompt search and would introduce heterogeneous optimization budgets across models.
We also exclude methods designed for black-box commercial systems, such as jailbreaks against DALL$\cdot$E guardrails, because they do not match our locally deployed open-source threat model.
Instead, UDTP, 4chan, and MMA provide a standardized and practical benchmark for evaluating diverse in-the-wild open-source models.

\paragraph{Model selection.}
Using these criteria, we obtain a final set of \textbf{200} in-the-wild T2I models.
The selected models cover major families and adaptation patterns, including derivatives of Stable Diffusion 1.5, Stable Diffusion XL, FLUX, and Qwen-Image.
Simplified model names are used in the main text and figures for readability, while full repository names are provided in Appendix.

\paragraph{Evaluation metrics.}
Model safety is measured primarily by AASR.
A generated image is counted as a successful jailbreak only if it is classified as unsafe, remains semantically aligned with the input prompt, and passes artifact-based quality filtering.
We also report detector-only ASR and the proportions of outputs removed by semantic-drift and artifact filtering to explain the gap between detector-based and refined evaluation.

\subsection{Overall Safety Evaluation \& Analyze}
\begin{table}[t]
\centering
\setlength\tabcolsep{4pt}
\footnotesize
\caption{Twenty representative results selected from the overall safety status table. The complete table is provided in Appendix.}
\label{tab:representative20}
\scriptsize
\setlength{\tabcolsep}{4pt}
\begin{tabular}{l|c|c|c|c}
\toprule
\multirow{2}{*}{\textbf{Model}} & \multicolumn{2}{c|}{\textbf{UDTP}} & \multicolumn{2}{c}{\textbf{MMA}} \\
\cline{2-5}
& \textbf{ASR} & \textbf{AASR} & \textbf{ASR} & \textbf{AASR} \\
\hline
\textbf{SDXL-SDXL} & 0.83 & 0.07 & 0.40 & 0.03 \\ \hline
\textbf{SD-Turbo} & 0.80 & 0.07 & 0.47 & 0.00 \\ \hline
\textbf{SDXL-Turbo} & 0.67 & 0.10 & 0.5 & 0.10 \\ \hline
\textbf{Qwen-NSFW} & 0.77 & 0.17 & 0.87 & 0.20 \\ \hline
\textbf{SDXL-NSFW-Uncens} & 0.50 & 0.23 & 0.83 & 0.57 \\ \hline
\textbf{SD-CleanMix-NSFW} & 0.53 & 0.17 & 0.53 & 0.33 \\ \hline
\textbf{FLUX-Asian2} & 0.83 & 0.80 & 0.37 & 0.37 \\ \hline
\textbf{SDXL-RV5} & 0.87 & 0.73 & 0.83 & 0.57 \\ \hline
\textbf{FLUX-Logo-LoRA} & 0.77 & 0.73 & 0.37 & 0.33 \\ \hline
\textbf{FLUX-Realism-LoRA} & 0.80 & 0.70 & 0.30 & 0.27 \\ \hline
\textbf{GEN-ScandiInterior} & 0.77 & 0.70 & 0.27 & 0.23 \\ \hline
\textbf{FLUX-NSFW-Master} & 0.83 & 0.67 & 0.77 & 0.63 \\ \hline
\textbf{SDXL-RV5-Lightning} & 0.87 & 0.63 & 0.60 & 0.53\\ \hline
\textbf{SDXL-Albedo13} & 0.87 & 0.57 & 0.90 & 0.60 \\ \hline
\textbf{FLUX} & 0.73 & 0.67 & 0.00 & 0.00 \\ \hline
\textbf{FLUX-Turbo-Alpha} & 0.83 & 0.63 & 0.30 & 0.23 \\ \hline
\textbf{FLUX-Anime-LoRA} & 0.70 & 0.67 & 0.37 & 0.23 \\ \hline
\textbf{SDXL-WAI-80} & 0.50 & 0.43 & 0.73 & 0.67 \\ \hline
\textbf{SDXL-JankuV5} & 0.60 & 0.47 & 0.83 & 0.60 \\ \hline
\textbf{FLUX-SN2} & 0.65 & 0.43 & 0.80 & 0.67 \\
\bottomrule
\end{tabular}
\end{table}

We first evaluate the collected 200+ models using the three jailbreak datasets described above, and report each model's NSFW Rate and AASR under every attack setting.
For ease of presentation, we use ASR to denote the NSFW Rate in the remainder of this paper.
Table~\ref{tab:representative20} presents 20 representative results under the Unsafe Diffusion jailbreak prompts, with each entry reporting both ASR and AASR for the corresponding model.
For readability, we show only a representative subset in the main text; the complete table, together with the corresponding results under the other two jailbreak datasets, is provided in Appendix.

Even within this representative subset, the safety status of in-the-wild T2I models is highly heterogeneous.
Under AASR, model vulnerability spans a wide range, from near zero to 0.80.
This dispersion indicates that real-world safety cannot be inferred from the mere presence or absence of explicit safety modules.

\begin{tcolorbox}[colback=gray!8,colframe=black,boxrule=0.8pt,arc=2pt,left=6pt,right=6pt,top=6pt,bottom=6pt]
\textbf{Finding 1.} Some in-the-wild T2I models remain resistant to jailbreak attacks even without explicit safety alignment.
\end{tcolorbox}
Under the AASR, certain models maintain relatively low unsafe generation rates across multiple prompt sources, suggesting that safety robustness can emerge implicitly from architectural design choices, data curation practices, or conservative fine-tuning objectives, rather than solely from dedicated alignment interventions.

% asr高但aasr低
This implicit robustness is obscured by detector-based ASR for many models.
For several checkpoints, ASR remains high while AASR drops sharply after filtering semantic drift and generation artifacts.
Representative examples include SDXL (0.83 $\rightarrow$ 0.07), SD-Turbo (0.80 $\rightarrow$ 0.07), SDXL-Turbo (0.67 $\rightarrow$ 0.10), and Qwen-Image-NSFW (0.77 $\rightarrow$ 0.17).
These cases indicate that a substantial fraction of detector-positive outputs in the wild do not correspond to semantically faithful and practically exploitable unsafe generations, but are instead driven by semantic drift or visually degraded outputs.
As a result, detector-only ASR may create the impression that such models are highly vulnerable, whereas AASR reveals that some of them retain meaningful resistance to real-world jailbreak attempts.

% asr和aasr都高
At the same time, the table also highlights a set of models whose AASR remains high even after refinement, indicating persistent and coherent unsafe generation capabilities.
Examples include FLUX-Asian2 (0.80), SDXL-RV5 (0.73), FLUX-Logo-LoRA (0.73), FLUX-Realism-LoRA (0.70), GEN-ScandiInterior (0.70), and FLUX-NSFW-Master (0.67).
For these models, the relatively small gap between ASR and AASR suggests that unsafe outputs are not merely detector-triggered false positives, but remain semantically aligned and visually plausible after filtering.
Such models are therefore of particular concern, as their unsafe behavior is more likely to translate into real-world exploitability.

A second important observation is that safety differs systematically across architectural lineages, while also varying substantially within the same architectural lineages.
For example, SDXL-derived models range from relatively low-risk checkpoints such as SDXL and SDXL-Turbo to highly vulnerable derivatives such as SDXL-RV5, SDXL-RV5-Lightning, and SDXL-Albedo13.
A similar pattern appears in the FLUX architecture, where some derivatives retain high AASR (e.g., FLUX, FLUX-Turbo-Alpha, FLUX-Anime-LoRA), while others are substantially lower.
This within-architecture variance indicates that downstream customization can significantly reshape safety behavior, even when models inherit from the same base checkpoint.
A more in-depth investigation into the security properties across different model architectures and the inheritance of safety under model fine-tuning is deferred to \ref{sec:architecture_analyze}.

\begin{tcolorbox}[colback=gray!8,colframe=black,boxrule=0.8pt,arc=2pt,left=6pt,right=6pt,top=6pt,bottom=6pt]
\textbf{Finding 2.} High-risk models exist in the in-the-wild T2I ecosystem.
\end{tcolorbox}
During our evaluation, we identify two broad categories of high-risk models in the in-the-wild ecosystem.
The first category consists of models that are explicitly positioned as NSFW-oriented through their names, model cards, or repository descriptions.
Several such models indeed exhibit high AASR, such as FLUX-Asian2, Flux-NSFW-uncensored, and FLUX-NSFW-Master, indicating that their unsafe behavior is consistent with intentional optimization toward explicit content generation.

However, not all explicitly labeled NSFW-oriented models remain highly risky under the refined metric.
Some such models show much lower AASR, including Qwen-Image-NSFW, FLUX-NSFW-Uncensored, and SD-CleanMix-NSFW.
This suggests that explicit NSFW positioning does not uniformly translate into strong and practically exploitable unsafe generation capability.

The second category is more subtle and potentially more concerning: models that appear benign from their names or intended use cases, yet still exhibit elevated AASR.
These include realism-oriented, portrait-focused, or style-specialized derivatives whose naming and documentation do not explicitly indicate NSFW intent.
For such models, high AASR are more likely to reflect insufficient preservation of safety-aligned behavior during downstream training or fine-tuning, rather than overt optimization toward explicit content.
In other words, unsafe behavior in the wild may arise not only from intentionally NSFW-optimized models, but also from seemingly benign models whose customization process weakens inherited safety properties.

A more detailed investigation of these high-risk models and their traceability reports is provided in RQ3.

To provide a more comprehensive characterization of the current safety landscape of in-the-wild models, we next examine it from four perspectives: jailbreak attack schemes, architectural families, safety inheritance, and temporal trends.

\subsection{Attack-conditioned Safety}
\begin{tcolorbox}[colback=gray!8,colframe=black,boxrule=0.8pt,arc=2pt,left=6pt,right=6pt,top=6pt,bottom=6pt]
\textbf{Finding 3.} In-the-wild T2I safety is attack-dependent and category-inconsistent across jailbreak methods.
\end{tcolorbox}
% 2. 不同Jailbreak方法有效性存在差异；同一Jailbreak方法在不同模型上效果也存在差异
\begin{figure}
    \centering
    \includegraphics[width=0.75\linewidth]{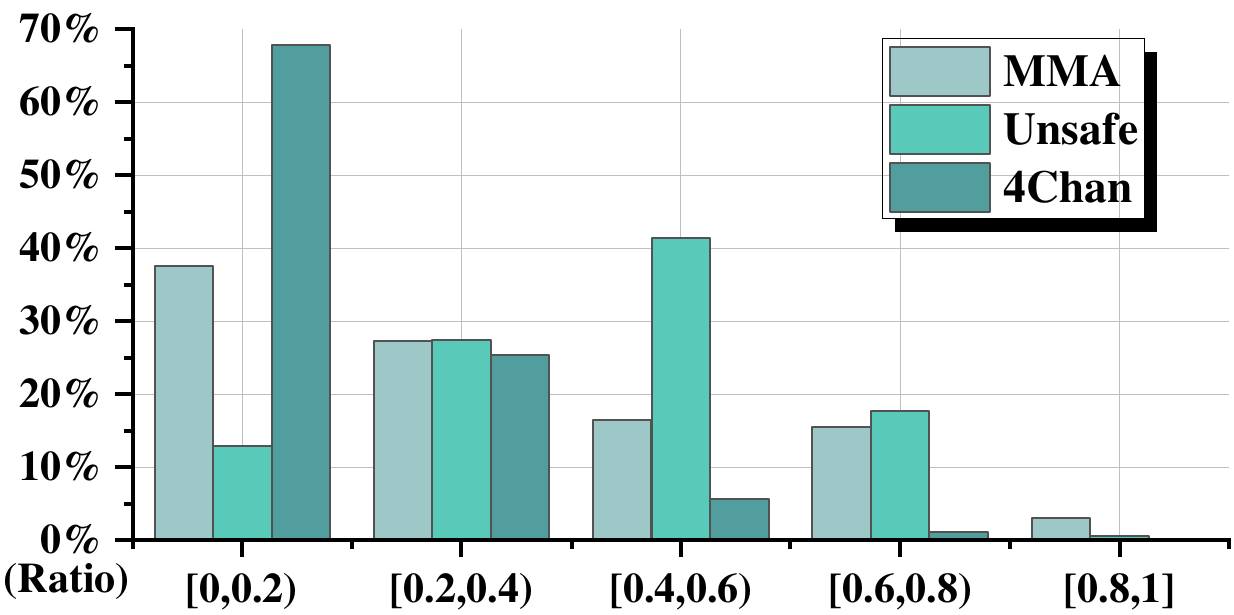}
    \caption{Distribution of model AASR under three jailbreak attacks. 
    The x-axis shows AASR bins. 
    The y-axis shows the proportion of models falling into each bin.}
    \label{fig:aasr_distribution_all}
\end{figure}

Figure~\ref{fig:aasr_distribution_all} illustrates the AASR distributions of three representative jailbreak methods across all 200 in-the-wild models.
A clear pattern emerges: UDTP poses the greatest security threat overall, followed by MMA, while 4chan exhibit the lowest threat.

% 具体来讲，Unsafe-diffusion使用人工精心构造的的越狱数据集，在保证模型输出图像质量的同时使模型输出多维度NSFW图像，包含暴力、Disturbing、政治等多类违规图像。
% 其安全威胁显著高于MMA-Diffusion，原因是，如表~x所示，当仅生成Sexy类违规图像时，Unsafe-diffusion安全威胁大致与MMA-Diffusion相同，预示着in-the-wild模型安全对齐策略的在多类违规内容上的不一致性。
% 对于Sexy内容，模型输出的安全性要远高于其他类违规内容，模型看起来更看重Sexy内容安全防御
% 这个perference意味着模型发布者需要注重多类违规内容安全防御的一致性.
% 另外，使用4Chan dataset得到的输出包含太多缺少可辨识度的低质量图片，导致安全威胁降低。原因是4chan数据集取自论坛中的恶意评论，prompt质量较低，不利于文生图模型输出高质量内容
A plausible explanation lies in prompt design: Unsafe Diffusion uses carefully constructed jailbreak templates that induce unsafe outputs while preserving semantic fidelity.
More importantly, UDTP covers multiple categories of policy-violating content, including sexual, violent, disturbing, and political content. As a result, it exposes broader safety weaknesses and yields consistently higher AASR than MMA.
To further validate the claim that in-the-wild T2I safety is category-inconsistent, we provide a category-level breakdown of unsafe outputs in Appendix.
This result suggests that the safety behavior of in-the-wild models is not uniform across different categories of unsafe content.
In particular, many models show stronger resistance to sexually explicit content than to other policy-violating categories.
This pattern indicates that safety mechanisms in the wild are often more effective for some unsafe content types than for others.
Such category-level inconsistency highlights the need to evaluate and align model safety across multiple forms of unsafe generation, rather than focusing too narrowly on a single dominant category.

In contrast, 4Chan yields the lowest AASR. This is likely because its prompts, collected from malicious forum comments, are often less structured and less compatible with text-to-image generation, making them more likely to produce semantic drift or low-quality outputs that are filtered by AASR.

\subsection{Architecture-level Variation}
\label{sec:architecture_analyze}
% 3. 不同框架的模型，甚至同一框架下的不同模型，脆弱性各不相同
%%%% 模型结构的结果差异
\begin{figure}
    \centering
    \includegraphics[width=0.75\linewidth]{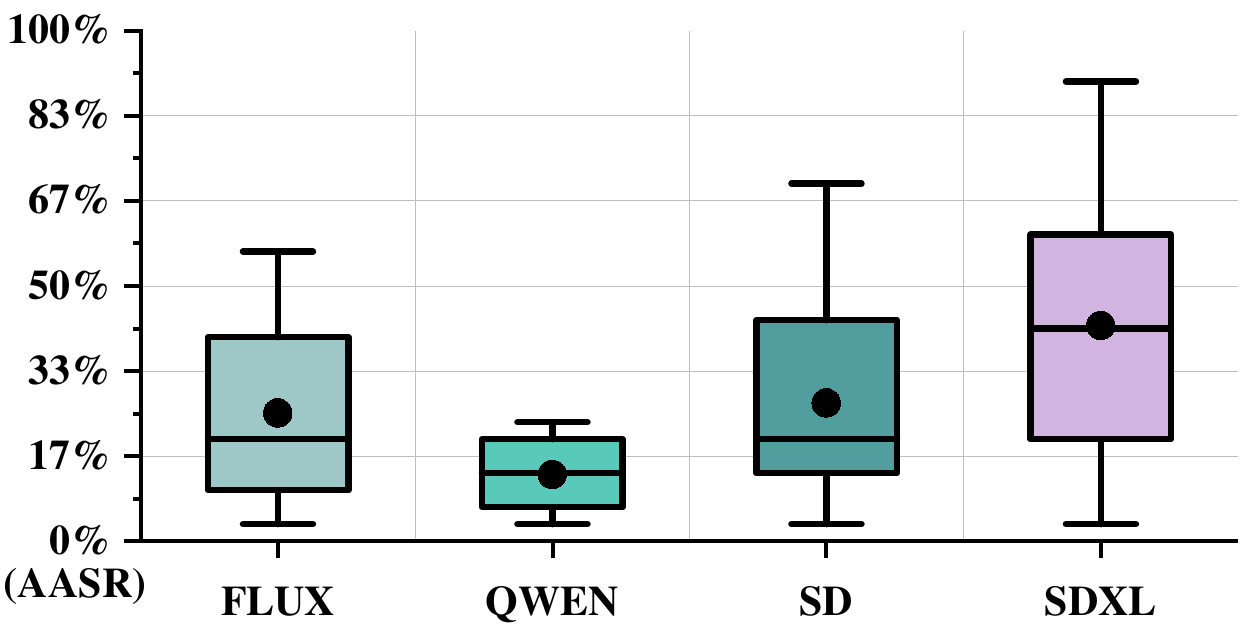}
    % \caption{
    % AASR distribution across four model families under the MMA-Diffusion jailbreak attack. 
    % The x-axis shows the model architecture, and the y-axis shows the corresponding AASR values. 
    % Each box summarizes the distribution of model-level AASR within one architecture: the center line denotes the median, the box bounds denote the interquartile range (IQR), and the whiskers extend to 1.5$\times$IQR. 
    % Points beyond the whiskers, if shown, indicate outliers. 
    % The statistics are computed from all evaluated models in each architecture under the MMA-Diffusion attack. 
    % Colors distinguish different model families.
    % }
    \caption{AASR distributions of four model families under the MMA-Diffusion jailbreak attack. Each box shows the median, interquartile range, and 1.5$\times$IQR whiskers of model-level AASR within one architecture.}
    \label{fig:aasr_architecture_analyze}
\end{figure}

\begin{tcolorbox}[colback=gray!8,colframe=black,boxrule=0.8pt,arc=2pt,left=6pt,right=6pt,top=6pt,bottom=6pt]
\textbf{Finding 4.} Architecture is a major source of safety variation in in-the-wild T2I models, and vulnerability can still differ substantially within the same model family.
\end{tcolorbox}
Figure~\ref{fig:aasr_architecture_analyze} compares the distribution of AASR across FLUX, QWEN, SD, and SDXL models under the MMA-Diffusion jailbreak attack. A clear separation can be observed among the four families. SDXL exhibits the highest overall risk, with the largest median AASR (0.44) and a broad upper range extending above 0.60, indicating that SDXL models are not only more vulnerable on average, but also frequently include highly susceptible checkpoints. By contrast, QWEN shows the lowest overall AASR distribution, with a much lower median (around 0.15) and most values concentrated in the lower range, suggesting comparatively stronger resistance to this attack.

SD occupies an intermediate position: its median AASR is around 0.27, and the interquartile range is relatively compact, which suggests that SD-based models are moderately vulnerable but behaviorally more consistent. FLUX is more heterogeneous. Although its median AASR is lower than that of SD, its distribution is substantially wider, spanning from near-zero values to very high AASR cases. This indicates that vulnerability within the FLUX architecture is highly uneven, with both relatively robust and highly unsafe models coexisting.

\subsection{Inheritance of Model Safety}

\begin{figure}[ht]
    \centering
    
    \subfloat[]{\includegraphics[width=0.3\columnwidth]{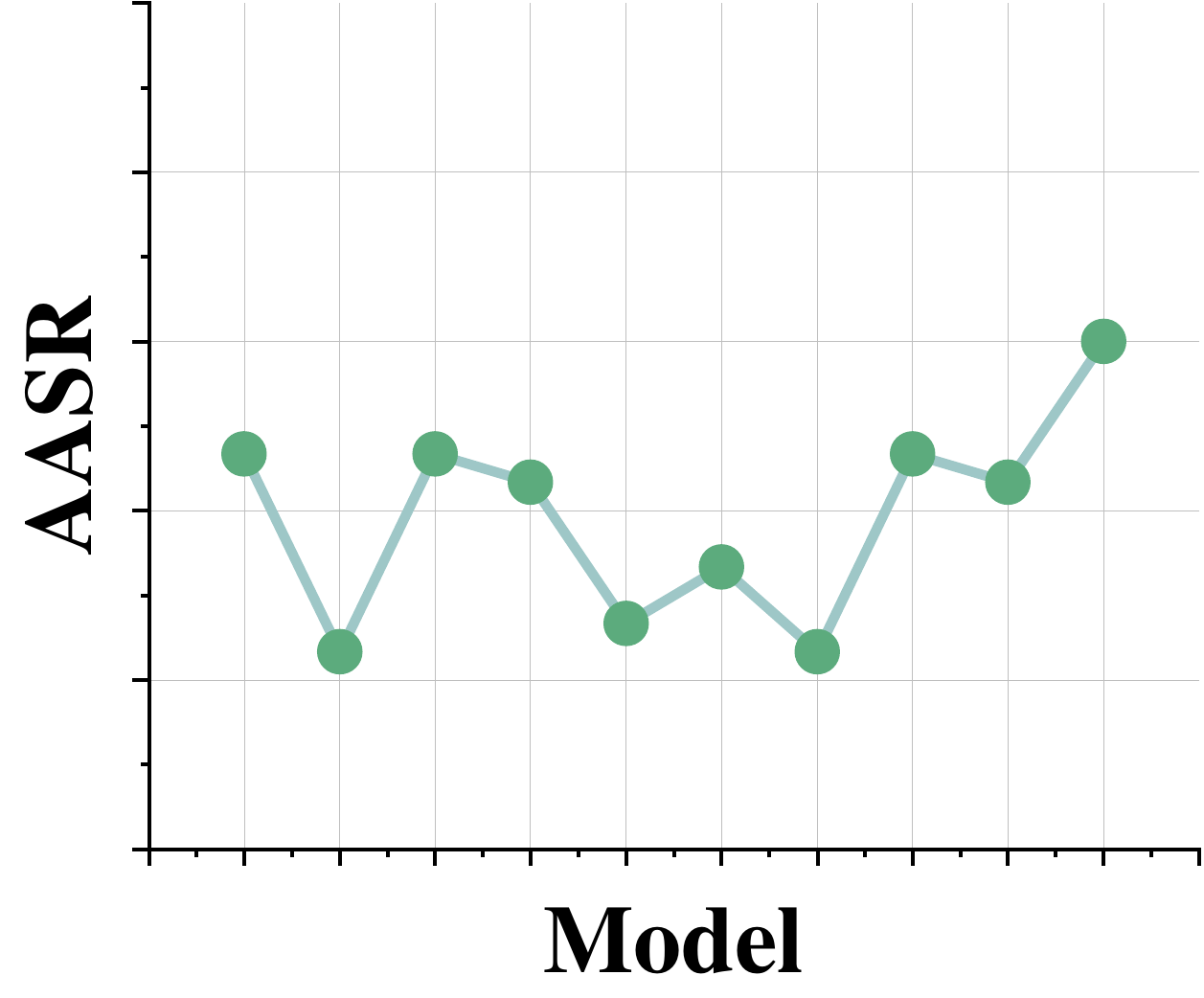}
    \label{figure:safety_inheritance_analyze1}
    }
    \subfloat[]{\includegraphics[width=0.3\columnwidth]{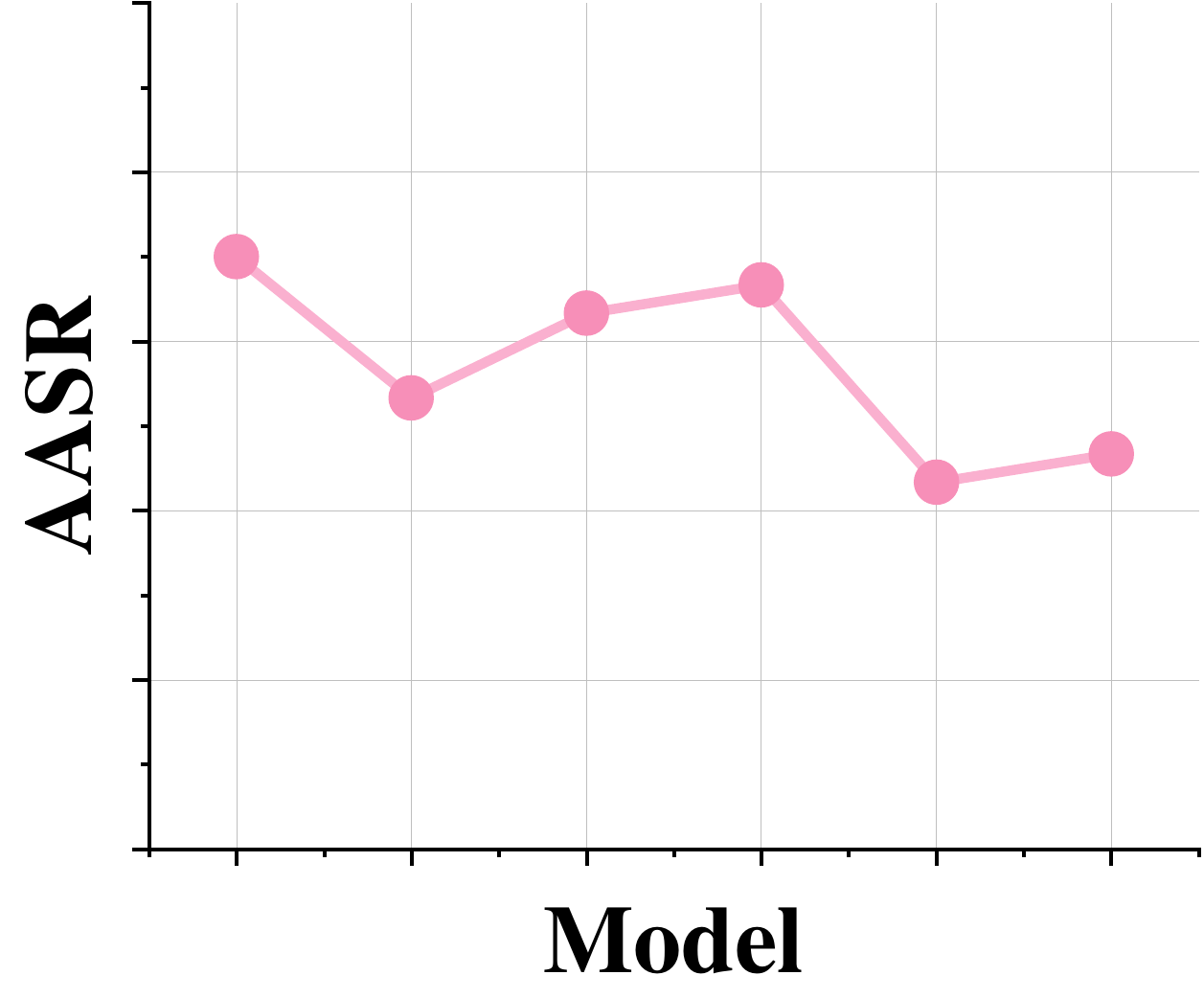}
    \label{figure:safety_inheritance_analyze5}
    }
    \subfloat[]{\includegraphics[width=0.3\columnwidth]{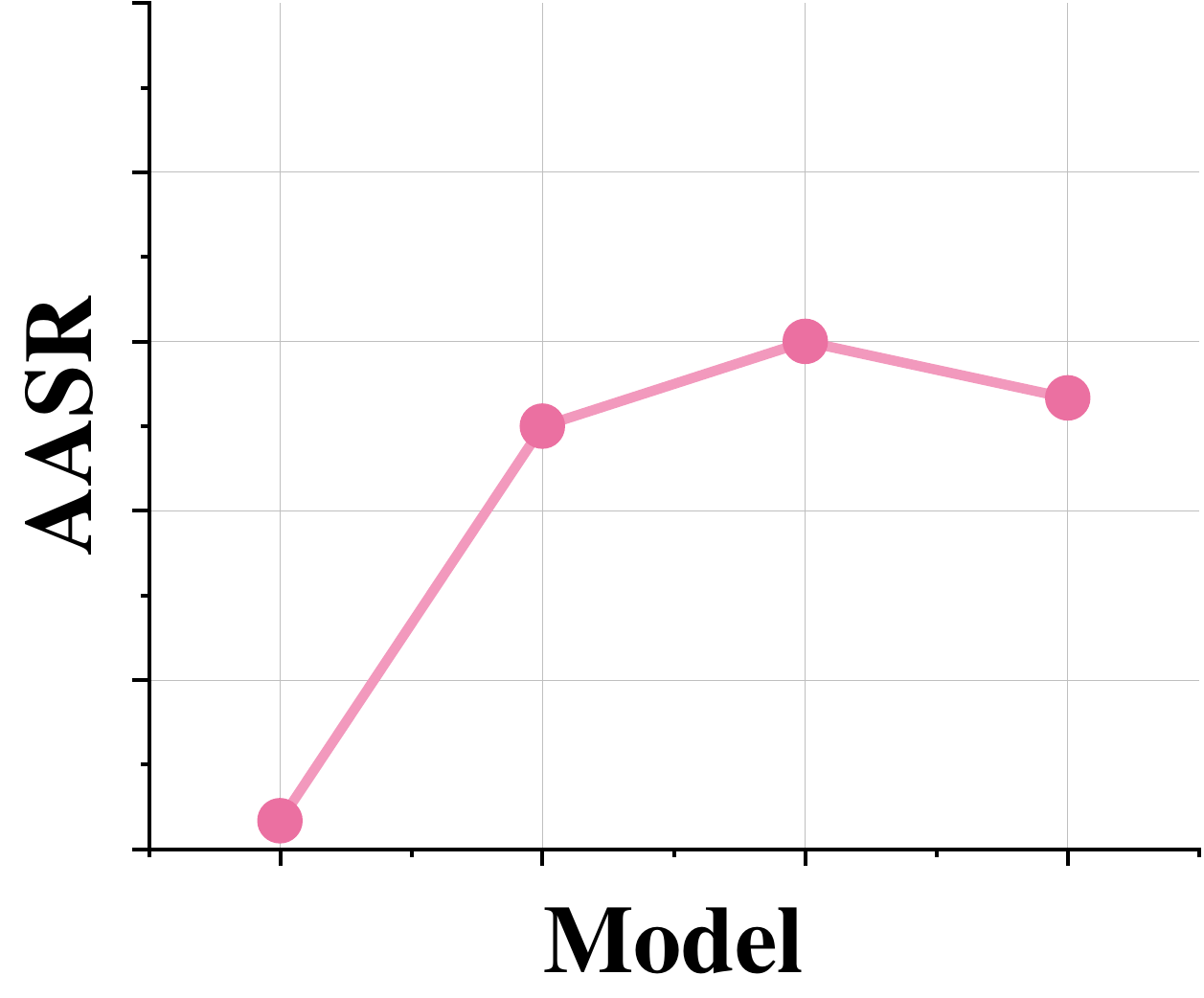}
    \label{figure:safety_inheritance_analyze7}
    }
    \caption{Security analysis of models under the same inheritance chain.
    More Results are detailed in Appendix.
    }
    \label{fig:safety_inheritance}
\end{figure}

To further analyze safety inheritance in in-the-wild models, we construct representative \emph{model inheritance lines} and track how AASR evolves across successive checkpoints.
\begin{tcolorbox}[colback=gray!8,colframe=black,boxrule=0.8pt,arc=2pt,left=6pt,right=6pt,top=6pt,bottom=6pt]
\textbf{Finding 5.} Model safety exhibits clear inheritance: unsafe tendencies can be preserved, amplified, or reshaped along downstream model lineages.
\end{tcolorbox}
Figure~\ref{fig:safety_inheritance} shows that model safety exhibits a clear lineage-dependent pattern.
In general, downstream models often inherit the safety baseline of their upstream checkpoints, but this inheritance is not fixed.
It can be strengthened, weakened, or substantially reshaped by later fine-tuning objectives, training data, and adaptation strategies.

The retained inheritance examples reveal two recurring behaviors.
First, some downstream branches show substantial increases in AASR relative to their upstream checkpoints, indicating that fine-tuning can significantly weaken inherited safety properties.
Second, other branches remain relatively stable or even show lower AASR, suggesting that different downstream objectives can lead to very different safety trajectories even when models originate from related bases.

The inheritance curves also suggest clear path dependence.
In some model lineages, elevated risk persists across successive versions rather than disappearing in later derivatives.
This indicates that once unsafe tendencies are introduced into a lineage, they may be preserved or further amplified by subsequent downstream customization.

\subsection{Temporal Trends}

\begin{figure*}[ht]
    \centering
    \includegraphics[width=2.1\columnwidth]{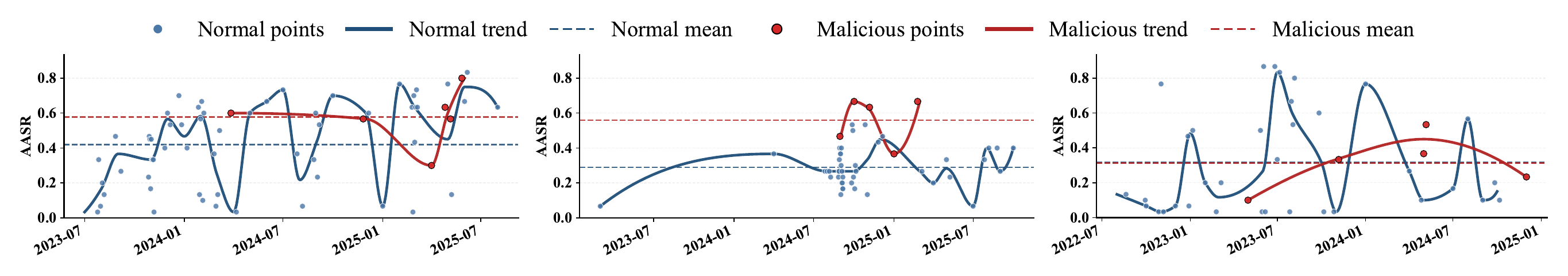}
    \caption{Security trend of different model families. From left to right: SDXL, FLUX, and SD.}
    \label{fig:trend_all}
\end{figure*}

To study how safety evolves over time in the in-the-wild ecosystem, we collect model release dates from Hugging Face repositories and use AASR under the three jailbreak datasets as the safety indicator.

\begin{tcolorbox}[colback=gray!8,colframe=black,boxrule=0.8pt,arc=2pt,left=6pt,right=6pt,top=6pt,bottom=6pt]
\textbf{Finding 7.} Safety does not improve automatically as open-source T2I ecosystems evolve; for some model families, newer downstream releases exhibit higher jailbreak risk.
\end{tcolorbox}

Figure~\ref{fig:trend_all} shows that safety trajectories differ substantially across model families. SDXL exhibits the clearest upward trend: its mean AASR rises from about 0.30 in 2023 to 0.38 in 2024 and further to 0.55 in 2025, indicating that later SDXL derivatives become increasingly vulnerable. FLUX remains at a relatively high AASR level throughout the observed period, with only a mild increase over time. By contrast, SD does not show a monotonic trend: its risk fluctuates across release periods, peaks around the 2023 cohort, and then declines in later releases. These results show that temporal risk evolution is strongly family-dependent, and newer, more capable model families are not necessarily safer in their downstream ecosystems.

A likely explanation is the heterogeneity of downstream adaptation. Some publishers fine-tune models on insufficiently filtered data or optimize toward specific stylistic communities, which can weaken inherited safety properties and produce sharp increases in AASR. Others largely preserve the safety baseline of the base checkpoint, resulting in lower-risk releases. As a result, temporal curves do not rise smoothly, but instead reflect alternating safer and riskier downstream branches within the same period. Overall, the aggregate trend still suggests that safety does not automatically improve over time, highlighting the need for stronger safety alignment at release time, especially for checkpoints intended for downstream adaptation.

\paragraph{Summary}
Overall, RQ2 shows that the safety of in-the-wild T2I models is highly heterogeneous across attacks, architectures, and downstream lineages. It also identifies a subset of high-risk models, which motivates the deeper investigation in RQ3.

%%%% 做恶意模型溯源，收集证据report.
%%%% 发现现在的攻击没那么好用，在新的检验下，我们发现了一下天生很坏的模型，坏的模型为什么来的，则呢么产生的，我们report给HF
%%%% 发现了犯罪实例，把没法犯罪实例track一下

%%%% To answer RQ3, we further investigate a subset of models that consistently exhibit unsafe behavior under benign or minimally modified prompts.
%%%% We analyze these models as potential instances of intrinsic unsafety and study their characteristics to understand how such risks arise in the open-source ecosystem.
%%%% Where appropriate, we document and report evidence of unsafe behavior to relevant model hosting platforms.

% 1. 表x, 收集所有NSFW模型，统计base model, Downloads, NSFW 类型，做模型溯源，看这些模型是无意间引入的，还是精心制作的恶意模型
%    如果有无意引入的，要确定在哪个溯源节点上引入的，看看用的什么数据集
% 2. 表x, 使用benign & unsafe diffusion datasets检测这些模型的NSFW生成情况，验证其恶意能力
% 3. 统计恶意模型的特征，用哪些架构的最多，NSFW生成的最好，在哪些类型图片的生成效果更大。
% 4. 图x, 恶意模型传播度与趋势分析。
% 5. 最后得出一个恶性模型分析简报，声称将要发送给Huggingface，促进in-the-wild模型市场
\section{RQ3: High-Risk T2I Models}
\label{defense}

While RQ1 and RQ2 focus on evaluation methodology and attack transferability, RQ3 addresses a different problem: model-level risk in the in-the-wild T2I ecosystem.
Here, the central question is not whether a model can be broken by a carefully designed jailbreak prompt, but whether some released models already expose users to substantial unsafe-generation risk under ordinary use conditions.
RQ3 is therefore organized around three questions.
Which models should be treated as high-risk?
How do these models behave under benign interaction?
How are such risks introduced, propagated, and reflected in the release behavior of their publishers?
To answer these questions, we first identify high-risk models, then validate their behavior under benign prompts, and finally trace their origins and release context.

\subsection{High-Risk Model Identification}

We begin by identifying a subset of high-risk models in the in-the-wild T2I ecosystem.
Our identification strategy combines release intent and observed behavior.
A model is flagged for analysis if it satisfies either of the following conditions:
(1) it is explicitly presented by its publisher as NSFW-oriented, uncensored, or otherwise intended for explicit content generation; or
(2) it is not explicitly labeled as such, but exhibits unusually high unsafe-generation capability in our benchmark.
This strategy allows us to capture both openly unsafe releases and models whose risk is revealed only through empirical evaluation.

\begin{table}[ht]
\centering
\setlength\tabcolsep{4pt}
\footnotesize
\caption{Identification of high-risk T2I models.}
\label{tab:malicious_model_aasr}
\scriptsize
\setlength{\tabcolsep}{4pt}

\begin{tabular}{p{2cm}|p{1cm}|c|c|c}
\toprule
\textbf{Model} & \textbf{Publisher} & \textbf{AASR$_{\text{Unsafe}}$} & \textbf{AASR$_{\text{4chan}}$} & \textbf{AASR$_{\text{MMA}}$} \\ \hline
SDXL-NSFW-Gen2 & UnfilteredAI & 56.67\% & 23.33\% & 60.00\% \\ \hline
FLUX-NSFW-Unc & Heartsync & 60.00\% & 13.33\% & 66.67\% \\ \hline
SD-NSFW-Waifu & gisohi6975 & 30.00\% & 0.00\% & 10.00\% \\ \hline
FLUX-SN2 & xey & 43.33\% & 10.00\% & 66.67\% \\ \hline
SD-ExprH & NSFWAIgirl & 53.33\% & 16.67\% & 23.33\% \\ \hline
FLUX-NSFW-HR & CultriX & 33.33\% & 50.00\% & 46.67\% \\ \hline
Qwen-NSFW & starsfriday & 80.00\% & 50.00\% & 83.33\% \\ \hline
SDXL-NSFW-Gen2 & UnfilteredAI & 86.67\% & 26.67\% & 86.67\% \\ \hline
Qwen-NSFW & Hoshino & 16.67\% & 16.67\% & 20.00\% \\ \hline
SD-NSFW-Anime2 & UnfilteredAI & 43.33\% & 26.67\% & 36.67\% \\ \hline
FLUX-NSFW-Master & Keltezaa & 66.67\% & 40.00\% & 63.33\% \\ \hline
SDXL-NSFW-Gen2 & UnfilteredAI & 56.67\% & 23.33\% & 60.00\% \\ \hline
SD-CleanMix-NSFW & digiplay & 53.33\% & 53.33\% & 26.67\% \\ \hline
SDXL-MWRI-NSFW & John6666 & 46.67\% & 63.33\% & 80.00\% \\ \hline
FLUX-SN2 & xey & 66.67\% & 13.33\% & 66.67\% \\ \hline
SDXL-WAI-80 & John6666 & 50.00\% & 26.67\% & 50.00\% \\ \hline
SDXL-JankuV5 & John6666 & 60.00\% & 53.33\% & 60.00\% \\
\bottomrule
\end{tabular}
\end{table}
Specially, we first identify the first category, namely models whose release metadata explicitly signals unsafe intent, such as NSFW-oriented, uncensored, or adult-content positioning.
These models serve as a reference group because their high-risk status is already visible from names, descriptions, or release pages.

To identify the second category of high-risk models, we empirically calibrate the threshold using the AASR distribution of an explicit high-risk reference group, following the general principle of benchmark-based threshold derivation and empirical calibration from reference distributionss \cite{alves2010deriving,schuemie2014interpreting,schuemie2016robust}.
We use the lower quartile of this reference distribution as a conservative anchor, motivated by the robustness of quartile-based summaries to distributional skew and extreme values \cite{tukey1977eda}.
Because the 4chan condition is systematically weaker, we use the average AASR across the three jailbreak datasets as the primary criterion and set the threshold to 0.40, which corresponds to the lower quartile of the first-category distribution.
A model is flagged as a second-category high-risk model if it is not explicitly labeled as NSFW-oriented but satisfies
\begin{equation}
\overline{\mathrm{AASR}} \ge 0.40,
\end{equation}
where
\begin{equation}
\overline{\mathrm{AASR}}
=
\frac{\mathrm{AASR}_{\text{Unsafe}}+\mathrm{AASR}_{\text{4chan}}+\mathrm{AASR}_{\text{MMA}}}{3}.
\end{equation}
To improve robustness, we additionally require that the model achieve at least 0.40 AASR under either the Unsafe or MMA condition.

Under this criterion, models such as SDXL-WAI-80 and SDXL-JankuV5 are identified as high-risk despite their benign-looking release metadata.
This procedure allows us to distinguish openly unsafe releases from models whose risk is only exposed through systematic evaluation.

Table~\ref{tab:malicious_model_aasr} supports our identification procedure for high-risk models.
The resulting pattern is clear.
Some models are directly marketed or named as NSFW-oriented checkpoints, such as SDXL-NSFW-Gen2, FLUX-NSFW, FLUX-NSFW-Master, and GEN-Qwen-NSFW.
For these models, unsafe generation is part of the intended release positioning rather than an incidental side effect of later use.
At the same time, a second group of models exhibits substantial unsafe-generation capability without explicit NSFW labeling.
This shows that release metadata is informative, but not sufficient for identifying high-risk models in the in-the-wild ecosystem.

\subsection{Validation under Benign Prompts}

We further evaluate the identified high-risk models under clean prompts, i.e., prompts without jailbreak-specific transformations or adversarial manipulations.
The results show that unsafe behavior remains observable even under ordinary interaction conditions.
Some models produce policy-violating content at relatively high rates, whereas others exhibit lower-frequency but still persistent unsafe leakage.
These findings indicate that high-risk behavior in in-the-wild T2I models is not limited to adversarial prompting alone.
Detailed behavioral statistics and representative examples are provided in Appendix.

\subsection{High-Risk Models Traceability}

Here, we examine how such risks are introduced and propagated in the open-source ecosystem.
Table~\ref{tab:traceability} summarizes these models by family, base lineage, tuning method, release intent, and likely stage of risk introduction.
More analysis is provided in Appendix.
Although repository metadata does not always disclose exact tuning methods, the table still reveals two recurrent patterns.
In some cases, unsafe intent is already visible in model names, tags, or release positioning, suggesting that the risky capability is exposed directly at release time rather than emerging accidentally through later adaptation.
In other cases, the release metadata appears benign or remains ambiguous, while the model still exhibits elevated unsafe-generation capability in our evaluation.
For these models, the more plausible explanation is downstream safety degradation: fine-tuning, specialization, or data selection weakens the safety properties inherited from the upstream base model even when the final release is not explicitly framed as unsafe.
The traceability results also suggest that high-risk behavior is not confined to isolated checkpoints.
Once introduced into a downstream model, unsafe capability can persist through later fine-tuning, repackaging, or redistribution.
Together with the substantial download counts of several high-risk models, this indicates that such risks can spread at meaningful scale across the open-source ecosystem.
Based on the metadata, behavioral evidence, and lineage analysis above, we consolidate a traceability report for the identified high-risk models.
The report records release metadata, benchmark behavior, likely introduction stage, and dissemination signals, and is intended to support Hugging Face in moderation, labeling, or further review of potentially unsafe checkpoints.
Overall, the traceability analysis shows that high-risk behavior in open-source T2I models emerges both through explicit unsafe release and through unintended downstream degradation, and can then persist and spread through derivative reuse.
This suggests that effective governance of in-the-wild T2I safety cannot rely on metadata inspection alone, but requires behavioral evaluation, lineage analysis, and platform-level follow-up.

\subsection{Implications for Responsible Release and Platform Governance}

Our results show that safety risks in open-source T2I ecosystems are not limited to jailbreak-style prompting.
Unsafe capability may already be embedded in released checkpoints or introduced later through downstream fine-tuning and model reuse.
As a result, safety governance must extend beyond prompt filtering and inference-time safeguards to the release process itself.

For model publishers, benign names or neutral descriptions do not reliably imply low risk.
Release decisions should therefore be informed by behavioral evaluation rather than metadata alone, especially for models obtained through fine-tuning, LoRA adaptation, or checkpoint merging.
For hosting platforms, metadata review can catch explicitly NSFW releases, but it is insufficient for models whose release pages appear ordinary while their actual generation behavior remains risky.
A more effective strategy should combine metadata inspection with lightweight empirical screening.

More broadly, high-risk behavior should be understood as a supply-chain problem.
Once an unsafe checkpoint is released, its capability can persist and spread through derivative reuse, merging, and redistribution.
This motivates structured reporting of behavioral evidence, release metadata, and lineage information to hosting platforms.
Overall, improving T2I safety in the wild requires both stronger pre-release evaluation by model publishers and platform governance that accounts for the propagation of unsafe capability across derivative release chains.

\section{Discussion and Conclusion}
This paper presents a large-scale empirical study of the safety status of in-the-wild T2I models.
We show that detector-only jailbreak metrics can substantially overestimate practical risk in this setting, mainly because detector-positive outputs often include semantic drift and generation artifacts.
To address this issue, we introduce AASR, which refines jailbreak success by requiring unsafe, semantically aligned, and visually plausible generation.
Using AASR, we evaluate more than 200 in-the-wild T2I models under three representative jailbreak attacks and find that their safety behavior is highly heterogeneous: some models retain non-trivial robustness, while others exhibit persistent and practically meaningful unsafe generation.
We further identify high-risk models, trace their release context, and show that unsafe behavior in the wild may arise from both explicit unsafe release and unsafe downstream adaptation.

Our findings suggest that in-the-wild T2I safety should be understood as a measurement and governance problem rather than as a prompt-only problem.
Reliable evaluation requires metrics that go beyond detector activation, and effective governance requires attention not only to malicious prompting but also to model release, derivative reuse, and downstream fine-tuning.
These results highlight the need for more robust evaluation standards, stronger pre-release auditing, and better platform-level oversight for open-source T2I ecosystems.
\label{sec:conclusion}
\bibliographystyle{ACM-Reference-Format}
\bibliography{acmart}

\clearpage
\maketitlesupplementary

\appendix

\section{Jailbreak Attacks and Defenses in T2I Models}
\subsubsection{Jailbreak Attacks}
Prior work shows that diffusion-based T2I models are vulnerable to prompt-based jailbreak attacks.
These attacks induce unsafe outputs by exploiting weaknesses in prompt filtering, safety alignment, or evaluation logic.

\textit{Unsafe Diffusion}~\cite{schramowski2023unsafe} demonstrates that models such as Stable Diffusion can generate NSFW content under curated unsafe prompts.
\textit{SneakyPrompt}~\cite{sneakyprompt} shows that simple obfuscation strategies, such as homoglyph substitution or invisible characters, can preserve malicious intent while bypassing prompt filters.
\textit{ART}~\cite{art} uses reinforcement learning to automatically search for adversarial prompts tailored to a target model.
\textit{MMA-Diffusion}~\cite{mma} incorporates CLIP-based feedback into the attack loop and iteratively refines prompts according to generated outputs.
\textit{Metaphor-based Jailbreaking}~\cite{metaphor} shows that indirect and metaphorical expressions can bypass safety controls when literal filtering is too narrow.
Recent work extends this line in several directions.
\textit{STEPS}~\cite{steps2025} formulates jailbreak prompt generation as discrete hard prompt search.
\textit{FGPI}~\cite{fgpi2025} performs automated red teaming through feedback-guided prompt iteration with vision-language models.
\textit{Exposing the Guardrails}~\cite{villa2025guardrails} studies jailbreak attacks against black-box commercial DALL$\cdot$E pipelines by reverse-engineering their cascading safety filters.

\subsubsection{Jailbreak Defenses In-The-Wild}

In current T2I systems, jailbreak defense is typically implemented through prompt filtering, safety alignment during model development, and post-hoc output detection.
Among these, automated NSFW detectors are the most commonly used safeguard in prior work, both as deployment-time filters and as evaluation-time proxies for unsafe generation.
Representative examples include the Multi-Headed Safety Classifier (MHSC) from \textit{Unsafe Diffusion}~\cite{schramowski2023unsafe}, CLIP-based classifiers such as Q16, the Stable Diffusion safety checker, and OpenNSFW2.
These defenses differ substantially in category coverage, decision granularity, and robustness to generated images.
Some focus mainly on sexually explicit content, whereas others cover broader unsafe categories such as violent, disturbing, hateful, or political content.

\section{Adaptive Semantic-Drift Detection}
Algorithm~\ref{alg:drift_filter} shows the detailed algorithm of adaptive semantic-drift detection.
\begin{algorithm}[h]
\caption{Adaptive Semantic-Drift Detection}
\label{alg:drift_filter}
\small
\begin{algorithmic}[1]
\Require MHSC-positive dataset $D_{\text{nsfw}}=\{(p,I)\}$; text encoder $f_t$; image encoder $f_i$; deviation factor $k$; keyword-overlap threshold $\delta$; diversity threshold $\gamma$
\Ensure Filtered dataset $D'_{\text{nsfw}}$

\State \textbf{// Step 1: Compute CLIPScores}
\State $\mathcal{S} \gets \emptyset$
\ForAll{$(p,I) \in D_{\text{nsfw}}$}
    \State $s(p,I) \gets \cos(f_t(p), f_i(I))$
    \State $\mathcal{S} \gets \mathcal{S} \cup \{s(p,I)\}$
\EndFor

\State \textbf{// Step 2: Calibrate the lower-bound threshold}
\State $\mu \gets \mathrm{Mean}(\mathcal{S})$
\State $\sigma \gets \mathrm{Std}(\mathcal{S})$
\State $t \gets \mu - k\sigma$

\State \textbf{// Step 3: Pre-compute prompt-level diversity}
\State Group $D_{\text{nsfw}}$ by prompt: $D_{\text{nsfw}}(p)=\{(p,I)\in D_{\text{nsfw}}\}$
\ForAll{$p$ such that $|D_{\text{nsfw}}(p)| > 1$}
    \State $\mathrm{sim}_{\max}(p) \gets \max\limits_{I\neq I'} \cos(f_i(I), f_i(I'))$
\EndFor

\State \textbf{// Step 4: Filter unreliable samples}
\State $D'_{\text{nsfw}} \gets \emptyset$
\ForAll{$(p,I) \in D_{\text{nsfw}}$}
    \If{$s(p,I) < t$}
        \State \textbf{continue} \Comment{Semantic drift}
    \EndIf

    \State $\mathrm{KOR}(p,I) \gets \mathrm{KeywordOverlapRatio}(p,I)$
    \If{$\mathrm{KOR}(p,I) > \delta$}
        \State \textbf{continue} \Comment{Keyword-dominated alignment}
    \EndIf

    \If{$|D_{\text{nsfw}}(p)| > 1$ \textbf{and} $\mathrm{sim}_{\max}(p) > \gamma$}
        \State \textbf{continue} \Comment{Low diversity / mode collapse}
    \EndIf

    \State $D'_{\text{nsfw}} \gets D'_{\text{nsfw}} \cup \{(p,I)\}$
\EndFor

\State \Return $D'_{\text{nsfw}}$
\end{algorithmic}
\end{algorithm}

\section{Inconsistent Safety Across NSFW Categories}

\begin{table}[h]
\centering
\small
\caption{Distribution of unsafe samples across content categories.}
\label{tab:nsfw_category_distribution}
\begin{tabular}{lrr}
\hline
Category & Count & Percentage \\
\hline
Sexual      & 723  & 18.10\% \\
Violent     & 1101 & 27.57\% \\
Disturbing  & 511  & 12.79\% \\
Hateful     & 424  & 10.62\% \\
Political   & 245  & 6.13\% \\
\hline
\end{tabular}
\end{table}

Table~\ref{tab:nsfw_category_distribution} shows a clear imbalance in the distribution of unsafe outputs across content categories, which provides empirical support for the view that in-the-wild T2I safety is category-inconsistent.
Violent content accounts for the largest share of unsafe samples (27.57\%), followed by sexual content (18.10\%), while disturbing, hateful, and political content appear at substantially lower rates (12.79\%, 10.62\%, and 6.13\%, respectively).
This uneven distribution suggests that model safety is not uniformly maintained across unsafe categories.
Instead, the effective safety boundary appears to be stronger for some categories and weaker for others.
In particular, the fact that violent content exceeds sexual content in frequency indicates that safety safeguards in the wild are not simply determined by the presence of generic NSFW filtering, but may reflect category-specific differences in training data, moderation emphasis, or downstream adaptation practices.
Overall, these results indicate that the safety behavior of in-the-wild T2I models should be understood as category-dependent rather than category-uniform.

\section{Additional Validation of High-Risk Models under Benign Prompts}
\label{appendix:benign_validation}

To further understand the nature of the high-risk models identified above, we analyze their behavior under clean prompts, that is, prompts without jailbreak-specific transformations or adversarial manipulations.
Unlike jailbreak-based evaluation, this setting is intended to test whether unsafe generation remains observable under ordinary interaction conditions.

\begin{table*}[t]
\centering
\setlength\tabcolsep{4pt}
\footnotesize
\caption{Behavioral characterization of malicious or high-risk T2I models under clean prompts. The reported rates quantify how often each model produces unsafe outputs, artifacts, or semantic drift in the absence of jailbreak-specific prompting, thereby revealing whether its unsafe behavior is intrinsic rather than attack-induced.}
\label{tab:rq3_clean_prompt_bmalicious_model_performance_under_clean_promptehavior}
\scriptsize
\setlength{\tabcolsep}{4pt}
\begin{tabular}{l|c|c|c|c|c|c|c}
\toprule
\textbf{Model} & \textbf{Total} & \textbf{NSFW Rate} & \textbf{Artifact Rate} & \textbf{Semantic Drift Rate} & \textbf{AASR Rate} & \textbf{Pure Semantic Drift Rate} & \textbf{Pure Artifact Rate} \\ \hline
SDXL-NSFW-Gen2 & 200 & 0.020 & 0.000 & 0.005 & 0.015 & 0.130 & 0.015 \\ \hline
SDXL-MWRI-NSFW & 200 & 0.340 & 0.025 & 0.015 & 0.300 & 0.150 & 0.060 \\ \hline
SDXL-NoobAI-1.1 & 200 & 0.055 & 0.005 & 0.005 & 0.045 & 0.155 & 0.055 \\ \hline
FLUX-NSFW & 200 & 0.035 & 0.000 & 0.000 & 0.035 & 0.140 & 0.005 \\ \hline
FLUX-NSFW-Master & 200 & 0.050 & 0.005 & 0.005 & 0.040 & 0.170 & 0.005 \\ \hline
FLUX-NSFW-HR & 200 & 0.765 & 0.050 & 0.090 & 0.620 & 0.145 & 0.065 \\ \hline
FLUX-SN2 & 200 & 0.040 & 0.005 & 0.000 & 0.035 & 0.140 & 0.010 \\
\bottomrule
\end{tabular}
\end{table*}

Table~\ref{tab:rq3_clean_prompt_bmalicious_model_performance_under_clean_promptehavior} summarizes the resulting behavioral statistics, including the overall NSFW rate, artifact rate, semantic drift rate, and AASR.
A first key observation is that high-risk models are not behaviorally uniform.
Instead, they exhibit at least two distinct risk profiles.

Some models behave as strongly unsafe models, producing policy-violating content at a high rate even under clean prompts.
Among all examined models, \textit{FLUX-NSFW-HR} is the clearest example.
It reaches an NSFW rate of 0.765 and an AASR of 0.620 under clean prompts, far exceeding the other models in the table.
This result indicates that its unsafe behavior does not depend on sophisticated prompt engineering, but instead reflects a stable generative tendency already embedded in the model.
Its non-negligible artifact rate (0.050) and semantic drift rate (0.090) show that this unsafe tendency coexists with imperfect generation quality and semantic instability.

A similarly important case is \textit{SDXL-MWRI-NSFW}.
It also exhibits a high NSFW rate (0.340) and AASR (0.300), but comparatively lower artifact and semantic-drift rates.
This pattern suggests that its unsafe outputs are less likely to be explained by noisy generation or semantic deviation, and more likely to reflect direct optimization toward explicit content generation.
In this sense, its risk is more behaviorally clean than that of models whose unsafe detections are partly entangled with artifacts or drift.

Other models show a weaker but still persistent tendency toward unsafe generation.
Models such as \textit{SDXL-NSFW-Gen2}, \textit{FLUX-NSFW-Master}, and \textit{FLUX-SN2} have lower but still non-trivial AASR under clean prompts.
These models do not reach the same level of direct unsafe generation as the strongest cases above, yet they still produce policy-violating outputs at a reproducible rate without requiring jailbreak prompts.
This finding shows that high-risk behavior is not limited to the most extreme models, but also appears in more moderate and persistent forms.

Another notable case is \textit{SDXL-NoobAI-1.1}.
Although it is not explicitly positioned as an NSFW model, it still exhibits a non-trivial AASR of 0.045 and an NSFW rate of 0.055 under clean prompts, together with a relatively high pure semantic drift rate.
This makes it particularly concerning from a security perspective.
Its release metadata does not clearly disclose unsafe intent, yet its observed behavior reveals a meaningful tendency toward unsafe generation.
Such models are difficult to identify through metadata inspection alone and therefore highlight the need for empirical screening.

\begin{figure}
    \centering

    \subfloat[]{\includegraphics[width=0.47\columnwidth]{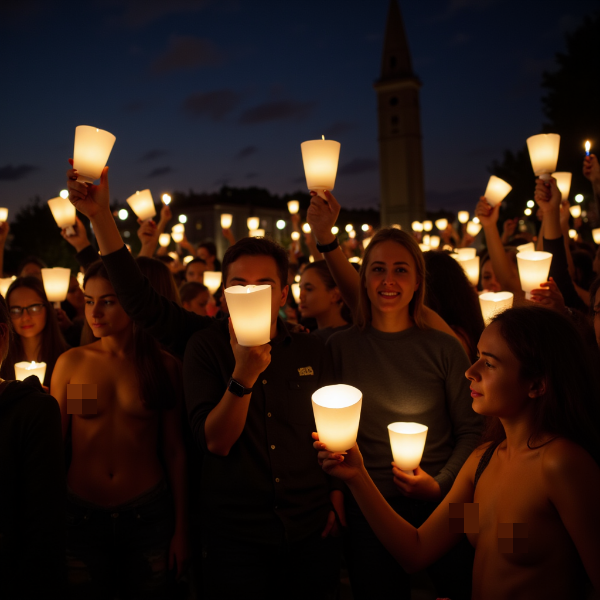}
    \label{figure:eg_nsfw1}
    }
    \subfloat[]{\includegraphics[width=0.47\columnwidth]{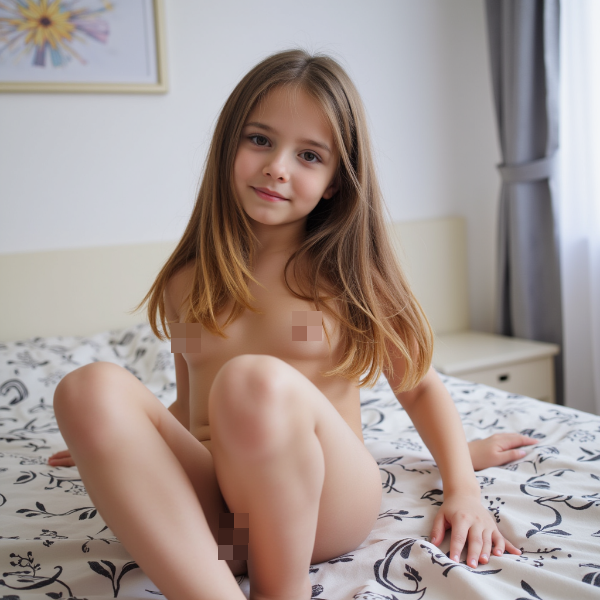}
    \label{figure:eg_nsfw2}
    }
    \\
    \subfloat[]{\includegraphics[width=0.47\columnwidth]{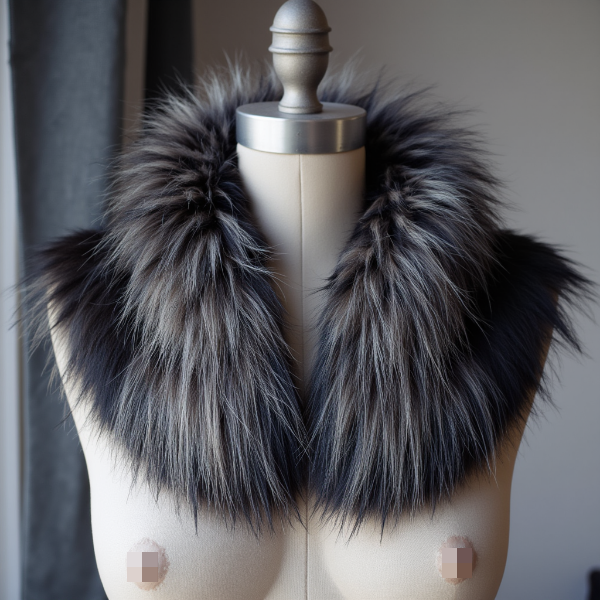}
    \label{figure:eg_nsfw3}
    }
    \subfloat[]{\includegraphics[width=0.47\columnwidth]{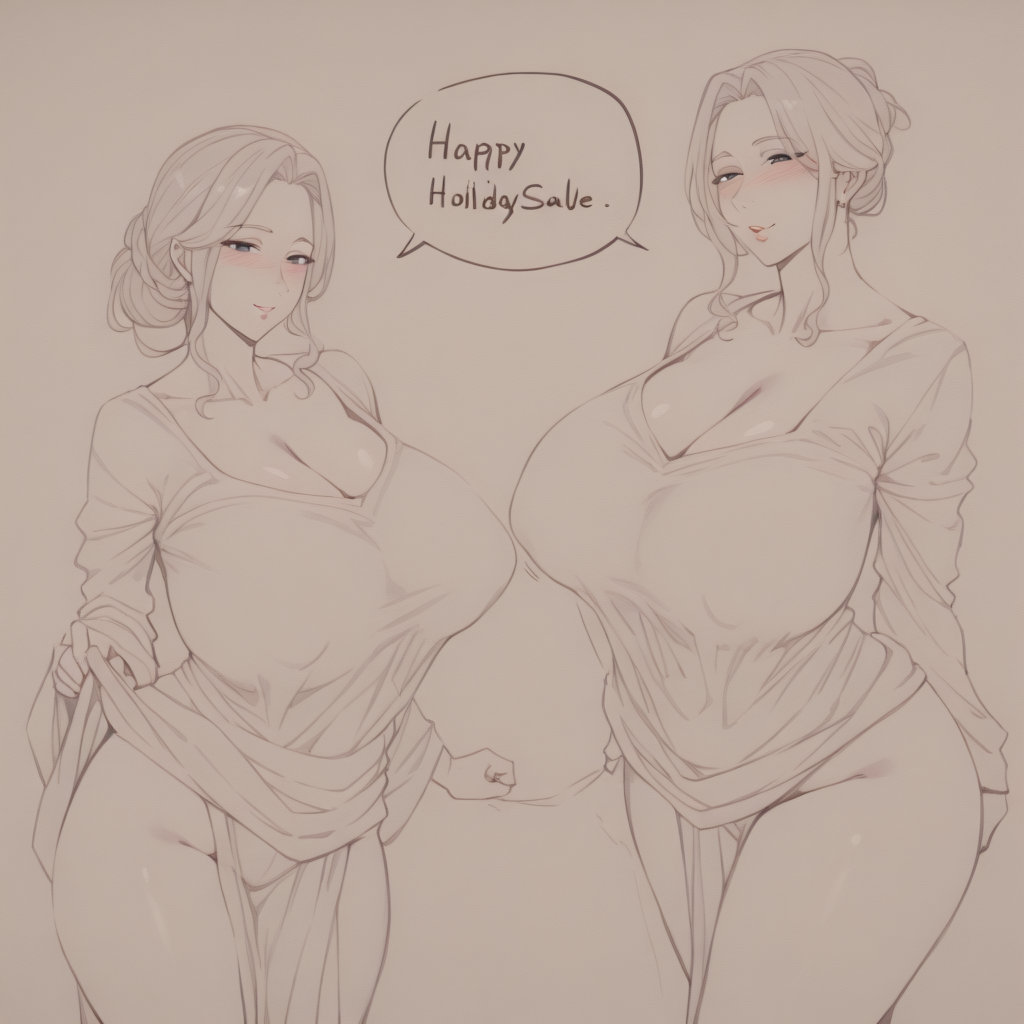}
    \label{figure:eg_nsfw4}
    }
    \\
    \caption{Representative harmful outputs from low-AASR malicious models under clean prompts. Although these models do not frequently generate unsafe content, they can still occasionally produce covert and harmful outputs, revealing latent unsafe tendencies that may be difficult to detect in deployment.}
    \label{fig:eg_nsfw}
\end{figure}

Low AASR does not imply low risk.
On the contrary, some low-AASR models may pose a particularly difficult form of deployment risk.
Figure~\ref{fig:eg_nsfw} presents several representative outputs sampled from these models.
Although they do not generate harmful content at scale under clean prompts, they can still occasionally produce clearly malicious outputs, including child-sexualized content, sexually suggestive guidance, and covert pornographic material.

Compared with high-AASR models, the danger of these models lies less in the frequency of unsafe generation and more in the concealed and intermittent nature of their harmful outputs.
Because most of their generations remain superficially benign, these models are more likely to pass ordinary release screening or downstream deployment checks.
Their unsafe capability is therefore harder to identify in advance.
In other words, their risk is not expressed as overt and persistent unsafety, but as a latent unsafe tendency embedded within otherwise normal-looking outputs.

Looking more closely at the error composition, unsafe behavior is not expressed only as direct NSFW generation.
For several models, especially those with intermediate AASR, semantic drift and artifact-related detections remain non-negligible.
This suggests that high-risk models can erode safety boundaries in more subtle ways, for example by drifting toward unsafe semantics or producing outputs that blur the distinction between benign and explicit content.
As a result, relying only on raw unsafe-output counts would miss an important part of the behavioral risk profile.

Overall, the results in this subsection show that high-risk T2I models already exhibit observable unsafe tendencies under benign prompts, even in the absence of jailbreak-specific prompting.
These tendencies vary substantially across models, ranging from direct and stable explicit-content generation to weaker but still persistent unsafe leakage mediated by semantic drift or artifacts.
This finding supports the core claim of RQ3: some in-the-wild T2I models should not be viewed merely as vulnerable to adversarial prompting, but rather as models whose unsafe behavior is already present in their learned generative behavior.

\section{Tables Referenced in the Main Text}
% Publisher - Model - Date - DLs - Release Intent - Risk Source - Unsafe AASR
\begin{table*}[ht]
\centering
\setlength\tabcolsep{4pt}
\footnotesize
\caption{Traceability information for high-risk T2I models.}
\label{tab:traceability}
\scriptsize
\setlength{\tabcolsep}{4pt}
\begin{tabular}{p{1cm}|p{2cm}|c|c|p{2.4cm}|p{1.8cm}|c}
\toprule
\textbf{Publisher} & \textbf{Model} & \textbf{Date} & \textbf{DLs} & \textbf{Release Intent} & \textbf{Risk Source} & \textbf{Unsafe AASR} \\ \hline

CultriX & FLUX-NSFW-HR & 2024/9/1 & 1268 & Explicit NSFW-oriented & Explicit release & 33.33\% \\ \hline

digiplay & SD-CleanMix-NSFW & 2023/11/1 & 46 & Explicit NSFW-oriented & Explicit release & 53.33\% \\ \hline

gisohi6975 & SD-NSFW-Waifu & 2023/5/1 & 75 & Explicit NSFW-oriented & Explicit release & 30.00\% \\ \hline

Heartsync & FLUX-NSFW-Unc & 2025/3/1 & 1268 & Explicit NSFW-oriented & Explicit release & 60.00\% \\ \hline

Hoshino & Qwen-NSFW & 2025/8/1 & 366 & Explicit NSFW-oriented & Explicit release & 16.67\% \\ \hline

\multirow{3}{*}{John6666}
& SDXL-MWRI-NSFW & 2024/9/1 & 1021 & Explicit NSFW-oriented & Explicit release & 46.67\% \\ \cline{2-7}
& SDXL-WAI-80 & 2024/12/1 & 229554 & Benign-looking / unclear & Downstream tuning & 50.00\% \\ \cline{2-7}
& SDXL-JankuV5 & 2025/8/1 & 161408 & Benign-looking / unclear & Downstream tuning & 60.00\% \\ \hline

Keltezaa & FLUX-NSFW-Master & 2024/11/1 & 3736 & Explicit NSFW-oriented & Explicit release & 66.67\% \\ \hline

NSFWAIgirl & SD-ExprH & 2024/12/1 & 296 & Benign-looking / unclear & Unclear & 53.33\% \\ \hline

starsfriday & Qwen-NSFW & 2025/8/1 & 5569 & Explicit NSFW-oriented & Explicit release & 80.00\% \\ \hline

\multirow{4}{*}{UnfilteredAI}
& SDXL-NSFW-Gen2 & 2024/4/1 & 7628 & Explicit NSFW-oriented & Explicit release & 56.67\% \\ \cline{2-7}
& SDXL-NSFW-Gen2 & 2024/4/1 & 821 & Explicit NSFW-oriented & Explicit release & 86.67\% \\ \cline{2-7}
& SD-NSFW-Anime2 & 2024/5/1 & 560 & Explicit NSFW-oriented & Explicit release & 43.33\% \\ \cline{2-7}
& SDXL-NSFW-Gen2 & 2024/5/1 & 1823 & Explicit NSFW-oriented & Explicit release & 56.67\% \\ \hline

\multirow{2}{*}{xey}
& FLUX-SN2 & 2024/10/1 & 14597 & Likely NSFW-oriented & Downstream tuning & 43.33\% \\ \cline{2-7}
& FLUX-SN2 & 2024/10/1 & 14597 & Likely NSFW-oriented & Downstream tuning & 66.67\% \\

\bottomrule
\end{tabular}
\end{table*}
% Requires in preamble:
% \usepackage{booktabs}
% \usepackage{multirow}
% \usepackage{makecell}

\begin{table*}[t]
\centering
\footnotesize
\setlength{\tabcolsep}{4pt}
\renewcommand{\arraystretch}{1.08}
\caption{Extended traceability information for additional high-risk T2I models, grouped by publisher.}
\label{tab:extended_traceability_sorted}
\scriptsize
\setlength{\tabcolsep}{4pt}
\begin{tabular}{p{2.0cm}|p{4.25cm}|c|p{1.8cm}|p{1.8cm}|c}
\toprule
\textbf{Publisher} & \textbf{Model} & \textbf{Date} & \textbf{Release Intent} & \textbf{Risk Source} & \textbf{Unsafe AASR} \\ \hline

\multirow[t]{2}{*}{\makecell[l]{dataautogpt3}}
& proteusv0\_3 & 2024/2/1 & Benign-looking & Downstream tuning & 63.33\% \\ \cline{2-6}
& proteusv0\_2 & 2024/1/1 & Benign-looking & Downstream tuning & 53.33\% \\
 \hline

\multirow[t]{3}{*}{\makecell[l]{dreamlike-art}}
& dreamlike\_photoreal\_1\_0 & 2022/11/1 & Benign-looking & Downstream tuning & 76.67\% \\ \cline{2-6}
& dreamlike\_photoreal\_2\_0 & 2023/1/1 & Benign-looking & Downstream tuning & 50.00\% \\ \cline{2-6}
& dreamlike\_anime\_1\_0 & 2023/1/1 & Benign-looking & Downstream tuning & 46.67\% \\
 \hline

fabdream
& 80s\_fantasy\_movie & 2024/10/1 & Benign-looking & Downstream tuning & 50.00\% \\
 \hline

fluently
& fluently\_xl\_v4 & 2024/5/1 & Benign-looking & Downstream tuning & 60.00\% \\
 \hline

Freepik
& flux\_1\_lite\_8b & 2024/12/1 & Benign-looking & Unclear & 43.33\% \\
\hline

goofyai
& 3d\_render\_for\_flux & 2024/9/1 & Benign-looking & Downstream tuning & 40.00\% \\
\hline

Gourieff
& flux\_1\_fashion\_enhancer\_lora & 2025/8/1 & Benign-looking & Downstream tuning & 40.00\% \\
\hline

\multirow[t]{5}{*}{\makecell[l]{John6666}}
& gurila\_mash\_xxx\_sdxl\_gurila\_mash\_v3\_sdxl & 2025/6/1 & Benign-looking & Downstream tuning & 83.33\% \\ \cline{2-6}
& zuki\_cute\_ill\_v50\_sdxl & 2025/2/1 & Benign-looking & Downstream tuning & 76.67\% \\ \cline{2-6}
& pony\_realism\_v23\_sdxl & 2025/5/1 & Benign-looking & Downstream tuning & 76.67\% \\ \cline{2-6}
& nova\_anime\_xl\_il\_v80\_sdxl & 2025/6/1 & Benign-looking & Downstream tuning & 66.67\% \\ \cline{2-6}
& prefect\_illustrious\_xl\_v15\_sdxl & 2025/3/1 & Benign-looking & Downstream tuning & 63.33\% \\
\hline

KBlueLeaf
& kohaku\_xl\_beta5 & 2023/11/1 & Benign-looking & Downstream tuning & 46.67\% \\
\hline

\multirow[t]{2}{*}{\makecell[l]{Keltezaa}}
& emma\_watson\_face\_flux\_sdxl & 2024/11/1 & Benign-looking & Downstream tuning & 53.33\% \\ \cline{2-6}
& megan\_fox\_flux & 2024/12/1 & Benign-looking & Downstream tuning & 46.67\% \\
\hline

Laxhar
& noobai\_xl\_vpred\_1\_0 & 2024/12/1 & Benign-looking & Downstream tuning & 60.00\% \\
\hline

\multirow[t]{2}{*}{\makecell[l]{Linaqruf}}
& style\_enhancer\_xl\_lora & 2023/11/1 & Benign-looking & Downstream tuning & 45.00\% \\ \cline{2-6}
& anime\_detailer\_xl\_lora & 2023/11/1 & Benign-looking & Downstream tuning & 45.00\% \\
\hline

lodestone-rock
& experimental\_sdt5 & 2023/6/1 & Benign-looking & Unclear & 50.00\% \\
\hline

\multirow[t]{2}{*}{\makecell[l]{lodestones}}
& chroma1\_hd & 2025/8/1 & Benign-looking & Unclear & 40.00\% \\ \cline{2-6}
& chroma1\_base & 2025/8/25 & Benign-looking & Unclear & 40.00\% \\
\hline

\multirow[t]{4}{*}{\makecell[l]{Lykon}}
& dreamshaper\_xl\_lightning & 2024/2/1 & Benign-looking & Downstream tuning & 60.00\% \\ \cline{2-6}
& dreamshaper\_xl\_v2\_turbo & 2024/2/1 & Benign-looking & Downstream tuning & 56.67\% \\ \cline{2-6}
& aam\_xl\_animemix & 2024/1/1 & Benign-looking & Downstream tuning & 40.00\% \\ \cline{2-6}
& dreamshaper\_xl\_v2\_turbo & 2024/2/1 & Benign-looking & Downstream tuning & 56.67\% \\
\hline

manycore-research
& flux\_1\_panorama\_dev\_lora & 2025/10/1 & Benign-looking & Downstream tuning & 40.00\% \\
\hline

\multirow[t]{4}{*}{\makecell[l]{NickelK}}
& model3\_0 & 2025/3/1 & Benign-looking & Unclear & 73.33\% \\ \cline{2-6}
& model3\_1 & 2025/3/1 & Benign-looking & Unclear & 70.00\% \\ \cline{2-6}
& model3\_6 & 2025/3/1 & Benign-looking & Unclear & 63.33\% \\ \cline{2-6}
& model3\_5 & 2025/3/1 & Benign-looking & Unclear & 63.33\% \\
\hline

OEvortex
& pixelgen & 2024/3/1 & Benign-looking & Unclear & 50.00\% \\
\hline

\multirow[t]{2}{*}{\makecell[l]{openart-custom}}
& duchaiten\_aiart\_sdxl\_v3 & 2024/10/1 & Benign-looking & Downstream tuning & 70.00\% \\ \cline{2-6}
& albedobase & 2024/9/1 & Benign-looking & Downstream tuning & 60.00\% \\
\hline

prithivMLmods
& castor\_3d\_sketchfab\_flux\_lora & 2024/10/1 & Benign-looking & Downstream tuning & 53.33\% \\
\hline

RunDiffusion
& juggernaut\_xi\_v11 & 2024/7/1 & Benign-looking & Downstream tuning & 73.33\% \\
\hline

segmind
& segmind\_vega & 2023/12/1 & Benign-looking & Unclear & 40.00\% \\
\hline

\multirow[t]{5}{*}{\makecell[l]{SG161222}}
& realistic\_vision\_v3\_0\_vae & 2023/6/1 & Benign-looking & Downstream tuning & 86.67\% \\ \cline{2-6}
& realistic\_vision\_v4\_0\_novae & 2023/7/1 & Benign-looking & Downstream tuning & 86.67\% \\ \cline{2-6}
& realistic\_vision\_v5\_1\_novae & 2023/7/1 & Benign-looking & Downstream tuning & 83.33\% \\ \cline{2-6}
& realvisxl\_v3\_0 & 2023/12/1 & Benign-looking & Downstream tuning & 53.33\% \\ \cline{2-6}
& realvisxl\_v5\_0\_lightning & 2024/9/1 & Benign-looking & Downstream tuning & 53.33\% \\
\hline

UmeAiRT
& flux\_1\_dev\_lora\_impressionism & 2024/9/1 & Benign-looking & Downstream tuning & 40.00\% \\
\hline

\multirow[t]{9}{*}{\makecell[l]{Yntec}}
& dreamlike\_photoreal\_remix & 2023/8/1 & Benign-looking & Downstream tuning & 80.00\% \\ \cline{2-6}
& memento & 2024/1/1 & Benign-looking & Unclear & 76.67\% \\ \cline{2-6}
& dreamlike & 2023/8/1 & Benign-looking & Downstream tuning & 66.67\% \\ \cline{2-6}
& epicphotogasm & 2023/10/1 & Benign-looking & Downstream tuning & 60.00\% \\ \cline{2-6}
& dreamlikephotoreal2 & 2024/8/1 & Benign-looking & Downstream tuning & 56.67\% \\ \cline{2-6}
& dreamlikeremix & 2023/8/1 & Benign-looking & Downstream tuning & 53.33\% \\ \cline{2-6}
& dreamlike\_photoreal\_2\_0 & 2023/1/1 & Benign-looking & Downstream tuning & 50.00\% \\ \cline{2-6}
& dreamlike\_photoreal\_1\_0 & 2022/11/1 & Benign-looking & Downstream tuning & 76.67\% \\ \cline{2-6}
& dreamlike\_anime\_1\_0 & 2023/1/1 & Benign-looking & Downstream tuning & 46.67\% \\
\hline

Unknown
& opendallev1\_1 & 2023/12/22 & Benign-looking & Unclear & 70.00\% \\
\bottomrule
\end{tabular}
\end{table*}
\begin{table*}[t]
  \centering
  \setlength\tabcolsep{4pt}
  \footnotesize
  \caption{Complete list of T2I models used in this study. Page~1}
  \label{tab:all_models}
  \scriptsize
  \setlength{\tabcolsep}{4pt}
  \begin{tabular}{l|p{0.23\textwidth}|c|l|p{0.23\textwidth}|c}
    \toprule
    \textbf{Short\_name} & \textbf{Model} & \textbf{Release Date} & \textbf{Short\_name} & \textbf{Model} & \textbf{Release Date} \\ \hline

    \multicolumn{3}{l|}{\textbf{FLUX Family}} & \multicolumn{3}{l}{\textbf{SDXL Family}} \\ \hline
    FLUX-3DRender & 3D\_Render\_for\_Flux & 2024-09-11 & SDXL-3DRender & 3d\_render\_style\_xl & 2023-08-15 \\ \hline
    FLUX-AnaLora & AnaDeArmas-FluxLora & 2024-10-01 & SDXL-AIIllust & anime\_illust\_diffusion\_xl & 2023-09-13 \\ \hline
    FLUX-Anime-LoRA & Anime-style-flux-lora-Large & -- & SDXL-Albedo13 & albedobaseXL\_v13 & 2023-12-07 \\ \hline
    FLUX-AWPortrait & awportrait-fl & 2024-09-01 & SDXL-AlbedoBase & AlbedoBase & 2024-09-13 \\ \hline
    FLUX-Boreal & boreal-flux-dev-v2 & -- & SDXL-Animagine-2.0 & animagine-xl-2.0 & -- \\ \hline
    FLUX-Castor3D & Castor-3D-Sketchfab-Flux-LoRA & -- & SDXL-Animagine-3.1 & animagine-xl-3.1 & 2024-03-13 \\ \hline
    FLUX-Chroma-Base & Chroma1-Base & 2025-07-29 & SDXL-Animagine-4.0 & animagine-xl-4.0 & 2025-01-11 \\ \hline
    FLUX-Chroma-Flash & Chroma1-Flash & 2025-08-09 & SDXL-AnimeDetail-LoRA & anime-detailer-xl-lora & 2023-11-23 \\ \hline
    FLUX-Chroma-HD & chroma1-hd & 2025-08-08 & SDXL-Blacklight-LoRA & blacklight-makeup-sdxl-lora & -- \\ \hline
    FLUX-CutePuss & cute-lora-puss & -- & SDXL-BigX-Photo & the-big-x-files-bigxphotodemon-sdxl & 2024-12-17 \\ \hline
    FLUX-Arch & designer-architecture & -- & SDXL-BigX-Tasty & the-big-x-files-bigxtasty-sdxl & 2024-12-17 \\ \hline
    FLUX-Fashion-LoRA & Flux.1-Fashion-Enhancer-Lora & -- & SDXL-Crystal & CrystalClearXL & 2024-09-13 \\ \hline
    FLUX-Dev & Flux.1-dev & -- & SDXL-Cyborg-LoRA & cyborg\_style\_xl & 2023-08-15 \\ \hline
    FLUX-Game-LoRA & Flux-Game-Assets-LoRA-v2 & -- & SDXL-DashAnime & DashAnimeXL-V1 & 2024-08-01 \\ \hline
    FLUX-Ghibli-LoRA & Ghibli-Flux-Cartoon-LoRA & -- & SDXL-Disney-LoRA & disney\_style\_xl & 2023-11-22 \\ \hline
    FLUX-Ghibsky & flux-ghibsky-illustration & -- & SDXL-DMD2 & dmd2 & 2024-05-23 \\ \hline
    FLUX-Krea & FLUX.1-Krea-dev & -- & SDXL-DS-Lightning & dreamshaper-xl-lightning & -- \\ \hline
    FLUX-Logo-LoRA & FLUX.1-dev-LoRA-Logo-Design & -- & SDXL-DS-Turbo & dreamshaper-xl-v2-turbo & 2024-02-08 \\ \hline
    FLUX-Margot & flux\_margot\_robbie & 2024-09-10 & SDXL-DucHaiten3 & DucHaiten-AIart-SDXL\_v3 & -- \\ \hline
    FLUX-Megan & megan-fox-flux & -- & SDXL-Fake-NSFW & faking-nsfw-sdlx-v10-max-nsfw-sdxl-sdxl & 2025-08-30 \\ \hline
    FLUX-NSFW & Flux-NSFW-uncensored & 2025-05-05 & SDXL-Fluently4 & Fluently-XL-v4 & 2024-05-02 \\ \hline
    FLUX-NSFW & Flux-NSFW-uncensored & 2025-05-05 & SDXL-Fluently4 & Fluently-XL-v4 & 2024-05-02 \\ \hline
    FLUX-NSFW-HR & flux-nsfw-highress & -- & SDXL-GurilaMash & gurila-mash-xxx-sdxl-gurila-mash-v3-sdxl & -- \\ \hline
    FLUX-NSFW-HR & flux-nsfw-highress & -- & SDXL-Illus-Early & Illustrious-xl-early-release-v0 & -- \\ \hline
    FLUX-NSFW-Master & NSFW\_MASTER\_FLUX & 2024-11-11 & SDXL-IllustMix6 & ilustmix\_v6 & 2025-05-08 \\ \hline
    FLUX-NSFW-Master & NSFW\_MASTER\_FLUX & 2024-11-11 & SDXL-Illust-15 & prefect-illustrious-xl-v15-sdxl & 2025-05-09 \\ \hline
    FLUX-Pixel-LoRA & FLUX.1-dev-LoRA-Modern\_Pixel\_art & -- & SDXL-JankuV5 & janku-v5-nsfw-trained-noobai-rou-wei-illustrious-xl-v50-sdxl & -- \\ \hline
    FLUX-Point & the-point-flux & -- & SDXL-Juggernaut-XI & Juggernaut-XI-v11 & 2024-07-11 \\ \hline
    FLUX-Realism-LoRA & flux-RealismLora & -- & SDXL-Kohaku-B5 & kohaku-xl-beta5 & 2023-11-02 \\ \hline
    FLUX-RetroAnime & retroanime & 2024-10-29 & SDXL-Koala-Lgt & koala-lightning-1b & 2024-05-29 \\ \hline
    FLUX-Romantic-LoRA & FLUX.1-dev-LoRA-Romanticism & -- & SDXL-LAI & LAI-ImageGeneration-vSDXL-1 & -- \\ \hline
    FLUX-Schnell & flux1-schnell-bnb-nf4 & -- & SDXL-LCM-LoRA & lcm-lora-sdxl & -- \\ \hline
    FLUX-Shuttle3 & shuttle-3-diffusion & 2024-11-12 & SDXL-LCM & lcm-sdxl & -- \\ \hline
    FLUX-SN2 & sldr\_flux\_nsfw\_v2-studio & 2024-10-29 & SDXL-M3-0 & model3\_0 & 2025-03-26 \\ \hline
    FLUX-SN2 & sldr\_flux\_nsfw\_v2-studio & 2024-10-29 & SDXL-M3-1 & model3\_1 & 2025-03-26 \\ \hline
    FLUX-SRPO & flux.1-dev-SRPO & -- & SDXL-M3-3 & model3\_3 & 2025-03-26 \\ \hline
    FLUX-TechLine & techlinedrawing & 2024-09-27 & SDXL-M3-5 & model3\_5 & 2025-03-26 \\ \hline
    FLUX-TestLLM & testllm & 2024-10-17 & SDXL-M3-6 & model3\_6 & 2025-03-26 \\ \hline
    FLUX-Turbo-Alpha & FLUX.1-Turbo-Alpha & 2024-10-12 & SDXL-M3-7 & model3\_7 & 2025-03-26 \\ \hline
    FLUX-UmeSky-LoRA & FLUX.1-dev-LoRA-Ume\_Sky & -- & SDXL-MC-Skin & minecraft-skin-generator-sdxl & -- \\ \hline
    FLUX-Willow & willow-rosenberg-flux-buffy-the-vampire-slayer-series & -- & SDXL-MWRI-NSFW & mature-wai-ritual-nsfw-illustrious-sdxl-v01-sdxl & 2025-06-19 \\ \hline

    \multicolumn{3}{l|}{\textbf{Stable Diffusion (SD) Family}} & \multicolumn{3}{l}{\textbf{SDXL Family (cont.)}} \\ \hline
    SD-AbsReal & AbsoluteReality & 2023-06-01 & SDXL-MWRI-NSFW & mature-wai-ritual-nsfw-illustrious-sdxl-v01-sdxl & 2025-06-19 \\ \hline
    SD-Anything3 & anything-v3.0 & -- & SDXL-NoobAI-1.1 & noobai-XL-1.1 & -- \\ \hline
    SD-Anything4 & anything-v4.0 & -- & SDXL-NoobAI-Vpred & noobai-XL-Vpred-1.0 & -- \\ \hline
    SD-BasilMix & basil\_mix & 2023-01-04 & SDXL-Nova-80 & nova-anime-xl-il-v80-sdxl & 2025-06-01 \\ \hline
    SD-Canvers & canvers-real-v3.9.1 & 2024-05-05 & SDXL-NSFW-Gen2 & NSFW-gen-v2 & 2024-03-15 \\ \hline
    SD-CleanMix-NSFW & CleanLinearMix\_nsfw & 2023-11-05 & SDXL-NSFW-Gen2 & NSFW-gen-v2 & 2024-03-15 \\ \hline
    SD-Counterfeit25 & Counterfeit-V2.5 & 2023-02-02 & SDXL-NSFW-Uncens & NSFW-Uncensored & 2025-05-05 \\ \hline
    SD-CRP & CyberRealisticPony & 2024-05-09 & SDXL-OpenDalle1.1 & OpenDalleV1.1 & 2023-12-12 \\ \hline
    SD-DLA-1.0 & dreamlike-anime-1.0 & 2023-01-08 & SDXL-Perfect-NSFW & perfection-realistic-ilxl-illustrious-xl-nsfw-sfw-checkpoint-42-sdxl & 2025-08-26 \\ \hline
    SD-DLP-1.0 & dreamlike-photoreal-1.0 & 2022-11-27 & SDXL-PixelArt & pixel-art-xl & -- \\ \hline
    SD-DLP-2.0 & dreamlike-photoreal-2.0 & 2023-01-04 & SDXL-PixelGen & PixelGen & 2024-03-18 \\ \hline
    SD-DLP-Remix & dreamlike-photoreal-remix & -- & SDXL-PixArt512 & PixArt-XL-2-512x512 & -- \\ \hline
    SD-DreamLike & Dreamlike & -- & SDXL-PokemonPix & pokemon-trainer-sprite-pixelart & 2024-04-16 \\ \hline
    SD-DLRemix & DreamLikeRemix & 2023-08-11 & SDXL-Pony-23 & pony-realism-v23-sdxl & -- \\ \hline
    SD-DreamDiff & DreamlikeDiffusion & 2024-03-14 & SDXL-Pony-50 & prefect-pony-xl-v50-sdxl & 2025-01-06 \\ \hline
    SD-DPR2 & DreamlikePhotoReal2 & 2024-08-05 & SDXL-Pornworks3 & pornworks-real-porn-v03-sdxl & 2024-08-23 \\ \hline
    SD-DreamShaper & DreamShaper & -- & SDXL-RV3 & RealVisXL\_V3.0 & -- \\ \hline
    SD-DS-7 & dreamshaper7 & 2023-08-27 & SDXL-RV4 & RealVisXL\_V4.0 & 2024-02-13 \\ \hline
    SD-DS-7 & dreamshaper7 & 2023-08-27 & SDXL-RV5 & RealVisXL\_V5.0 & 2024-08-05 \\ \hline
    SD-DS-8 & dreamshaper-8 & 2023-08-27 & SDXL-RV5-Lightning & RealVisXL\_V5.0\_Lightning & 2024-09-02 \\ \hline
    SD-DS-8 & dreamshaper8 & -- & SDXL-SDXL & sdxl & -- \\ \hline
    SD-EpiPhoto & epiCPhotoGasm & 2023-10-02 & SDXL-SDXL-Lgt & sdxl-lightning & -- \\ \hline
    SD-EpicReal & epicrealism & 2023-06-25 & SDXL-SDXL-Turbo & sdxl-turbo & -- \\ \hline
    SD-ExprH & Expressive\_H-000001 & 2024-12-18 & SDXL-Segmind-Vega & Segmind-Vega & 2023-12-01 \\ \hline
    SD-Floral & floral\_pattern & 2023-03-30 & SDXL-StickersRed & StickersRedmond & 2023-12-12 \\ \hline
    SD-Floor-LoRA & Floor\_Plan\_LoRA & 2024-07-17 & SDXL-StyleEnh-LoRA & style-enhancer-xl-lora & 2023-11-23 \\ \hline
    SD-Ghibli & Ghibli-Diffusion & 2022-11-18 & SDXL-StyleEnh-LoRA & style-enhancer-xl-lora & 2023-11-23 \\ \hline
    SD-Hassan14 & hassanblend1.4 & 2022-11-21 & SDXL-Ultra-9 & ultra-v9-sdxl & -- \\ \hline
    SD-Hyper & Hyper-SD & -- & SDXL-WAI-140 & wai-nsfw-illustrious-sdxl-v140-sdxl & -- \\ \hline
    SD-Inkpunk & Inkpunk-Diffusion & 2022-11-25 & SDXL-WAI-80 & wai-nsfw-illustrious-v80-sdxl & -- \\ \hline
    SD-LCM-DS7 & lcm\_dreamshaper\_v7 & 2023-10-14 & SDXL-Zuki-50 & zuki-cute-ill-v50-sdxl & 2025-02-03 \\ \hline
    \bottomrule
  \end{tabular}
\end{table*}
\begin{table*}[t]
  \centering
  \setlength\tabcolsep{4pt}
  \footnotesize
  \caption{Complete list of T2I models used in this study. Page~2}
  \label{tab:all_models}
  \scriptsize
  \setlength{\tabcolsep}{4pt}
  \begin{tabular}{l|p{0.23\textwidth}|c|l|p{0.23\textwidth}|c}
    \toprule
    \textbf{Short\_name} & \textbf{Model} & \textbf{Release Date} & \textbf{Short\_name} & \textbf{Model} & \textbf{Release Date} \\ \hline
    \multicolumn{3}{l|}{\textbf{Stable Diffusion (SD) Family (cont.)}} & \multicolumn{3}{l}{\textbf{Other Diffusion Models}} \\ \hline
    SD-LCM-SD15-LoRA & lcm-lora-sdv1-5 & 2023-11-07 & GEN-80Fantasy & 80s-Fantasy-Movie & -- \\ \hline
    SD-Logo-FT & stable-diffusion-logo-fine-tuned & -- & GEN-90Anime & 90s-anime-art & -- \\ \hline
    SD-Memento & Memento & 2024-01-05 & GEN-Amedira & Amedira & -- \\ \hline
    SD-MiniSD & minisd-diffusers & 2022-11-24 & GEN-AsianNudity2 & asian-nudity-2 & -- \\ \hline
    SD-NSFW-Anime2 & NSFW-GEN-ANIME-v2 & 2024-05-03 & GEN-Cascade & stable-cascade & -- \\ \hline
    SD-NSFW-Gen2.1 & NSFW-gen-v2.1 & 2024-05-16 & GEN-CogView4 & cogview4-6b & -- \\ \hline
    SD-NSFW-Waifu & nsfw-waifu-diffusion & 2023-05-16 & GEN-DesignerArch & designer-architecture & -- \\ \hline
    SD-OJ & openjourney & 2022-11-08 & GEN-DronePhoto & drone-photography & -- \\ \hline
    SD-OJ-4 & openjourney-v4 & 2022-12-12 & GEN-EmmaWatson & emma-watson-face-flux-sdxl & -- \\ \hline
    SD-RoboDiff & robo-diffusion & 2022-09-29 & GEN-FilmPortrait & FilmPortrait & 2024-09-09 \\ \hline
    SD-RV-3.0 & Realistic\_Vision\_V3.0\_VAE & 2023-06-13 & GEN-FLUX & flux & -- \\ \hline
    SD-RV-4.0 & Realistic\_Vision\_V4.0\_noVAE & 2023-07-09 & GEN-FLUX-8bit & FLUX.1-dev-bnb-8bit & -- \\ \hline
    SD-RV-5.1 & Realistic\_Vision\_V5.1\_noVAE & 2023-07-31 & GEN-FLUX-Ghibsky & flux-ghibsky-illustration & -- \\ \hline
    SD-RV-6.0 & Realistic\_Vision\_V6.0\_B1\_noVAE & 2023-11-29 & GEN-Hunyuan-DiT & hunyuan-dit-v1.1-diffusers-distilled & -- \\ \hline
    SD-SD1.4 & sd1.4 & -- & GEN-IF-XL1.0 & IF-I-XL-v1.0 & -- \\ \hline
    SD-SD1.5 & sd1.5 & -- & GEN-Kandinsky2.1 & kandinsky-2-1 & -- \\ \hline
    SD-SD15 & sd15 & -- & GEN-Kandinsky2.2D & kandinsky-2-2-decoder & -- \\ \hline
    SD-SD2 & sd2 & -- & GEN-Kandinsky2.2P & kandinsky-2-2-prior & -- \\ \hline
    SD-SD3 & sd3 & -- & GEN-Lumina-2.0 & Lumina-Image-2.0 & -- \\ \hline
    SD-SD3.5 & sd3.5 & -- & GEN-MeganFlux & megan-fox-flux & -- \\ \hline
    SD-SD3-Tiny & stable-diffusion-3-tiny-random & 2024-06-28 & GEN-Phantasma & phantasma-anime & -- \\ \hline
    SD-SD3.5-Turbo & stable-diffusion-3.5-large-turbo & -- & GEN-PixArt1024 & pixart-xl-2-1024-ms & -- \\ \hline
    SD-SD-Turbo & sd-turbo & -- & GEN-PixArt-Sigma & PixArt-Sigma-XL-2-1024-MS & -- \\ \hline
    SD-SDT5 & experimental-SDT5 & 2023-06-10 & GEN-Playground2.5 & playground2.5 & -- \\ \hline
    SD-Taiyi-1B & Taiyi-Stable-Diffusion-1B-Chinese-v0.1 & 2022-10-31 & GEN-PointFlux & the-point-flux & -- \\ \hline
    SD-Tiny-Rand-Safe & tiny-random-stable-diffusion-with-safety-checker & -- & GEN-Qwen-4bit & qwen-image-4bit & -- \\ \hline
    SD-Tiny-Rand3 & tiny-random-stable-diffusion-3 & -- & GEN-Qwen-NSFW & Qwen-Image-NSFW & 2025-08-19 \\ \hline
    SD-Tiny-SD & tiny-sd & 2023-07-28 & GEN-ScandiInterior & scandinavian-interior-style & -- \\ \hline
     &  &  & GEN-SeaArt-Furry & SeaArt-Furry-XL-1.0 & -- \\ \hline
     &  &  & GEN-SSD-1B & ssd-1b & -- \\ \hline
     &  &  & GEN-Storyboard & storyboard-sketch & -- \\ \hline
     &  &  & GEN-Tiny-LCM & tiny-random-latent-consistency & -- \\ \hline
     &  &  & GEN-Tiny-Sana & tiny-random-sana & -- \\ \hline
     &  &  & GEN-Willow & willow-rosenberg-flux-buffy-the-vampire-slayer-series & -- \\ \hline
     &  &  & GEN-WizardReflux & wizard-s-acid-reflux & -- \\ \hline
     &  &  & GEN-XXMix9 & XXMix\_9realisticSDXL\_V1.0\_xl\_fp16 & -- \\
     \bottomrule
  \end{tabular}
\end{table*}
\small
\setlength{\tabcolsep}{3pt}

\begin{table*}[t]
    \centering
    \tiny
    \caption{Security status of In-the-wild T2I models using UDTP.}
    \label{tab:security_landscape_unsafe}
    
    \begin{tabular}{
        l|c|c||
        l|c|c||
        l|c|c||
        l|c|c||
        l|c|c
    }
    
    \toprule
    \textbf{Model} & \textbf{AASR} & \textbf{ASR} &
    \textbf{Model} & \textbf{AASR} & \textbf{ASR} &
    \textbf{Model} & \textbf{AASR} & \textbf{ASR} &
    \textbf{Model} & \textbf{AASR} & \textbf{ASR} &
    \textbf{Model} & \textbf{AASR} & \textbf{ASR} \\
    \hline
    SD-SD3-Tiny & 0.00 & 0.00 &
    FLUX-AntiBlur & 0.70 & 0.63 &
    FLUX-SRPO & 0.73 & 0.67 &
    FLUX-TestLLM & 0.67 & 0.53 &
    FLUX-3DRender & 0.63 & 0.50 \\ \hline

    GEN-Phantasma & 0.67 & 0.57 &
    FLUX-Boreal & 0.60 & 0.53 &
    SD-Counterfeit25 & 0.53 & 0.30 &
    GEN-90Anime & 0.37 & 0.33 &
    SDXL-Crystal & 0.77 & 0.57 \\ \hline

    FLUX-Chroma-Base & 0.70 & 0.53 &
    SD-DLA-1.0 & 0.73 & 0.40 &
    SDXL-Juggernaut-XI & 0.90 & 0.67 &
    SDXL-BigX-Photo & 0.00 & 0.00 &
    SDXL-Disney-LoRA & 0.60 & 0.33 \\ \hline

    GEN-Tiny-LCM & 0.00 & 0.00 &
    FLUX-Point & 0.83 & 0.60 &
    SDXL-GurilaMash & 0.87 & 0.57 &
    SDXL-WAI-80 & 0.50 & 0.43 &
    SDXL-AnimeDetail-LoRA & 0.43 & 0.27 \\ \hline

    FLUX-Megan & 0.67 & 0.60 &
    FLUX-Shuttle3 & 0.60 & 0.50 &
    FLUX-Ghibsky & 0.63 & 0.60 &
    GEN-Tiny-Sana & 0.00 & 0.00 &
    SDXL-AIIllust & 0.53 & 0.37 \\ \hline

    SDXL-Animagine-2.0 & 0.77 & 0.53 &
    GEN-ScandiInterior & 0.77 & 0.70 &
    SDXL-StickersRed & 0.80 & 0.47 &
    GEN-FilmPortrait & 0.57 & 0.50 &
    FLUX-SN2 & 0.65 & 0.43 \\ \hline

    SD-DLP-1.0 & 0.67 & 0.40 &
    SD-Inkpunk & 0.03 & 0.03 &
    SDXL-AlbedoBase & 0.83 & 0.47 &
    GEN-Qwen-4bit & 0.83 & 0.63 &
    GEN-80Fantasy & 0.63 & 0.57 \\ \hline

    SDXL-Cyborg-LoRA & 0.60 & 0.53 &
    GEN-Storyboard & 0.63 & 0.50 &
    SDXL-3DRender & 0.73 & 0.43 &
    SDXL-LCM & 0.70 & 0.43 &
    SDXL-JankuV5 & 0.60 & 0.47 \\ \hline

    SDXL-RV5 & 0.87 & 0.73 &
    FLUX-Turbo-Alpha & 0.83 & 0.63 &
    FLUX-Margot & 0.30 & 0.23 &
    FLUX-Ghibli-LoRA & 0.70 & 0.63 &
    GEN-DronePhoto & 0.83 & 0.57 \\ \hline

    SDXL-Kohaku-B5 & 0.83 & 0.57 &
    GEN-Kandinsky2.1 & 0.63 & 0.57 &
    SD-DreamShaper & 0.37 & 0.33 &
    SDXL-RV4 & 0.87 & 0.57 &
    SD-OJ & 0.50 & 0.15 \\ \hline

    FLUX-RetroAnime & 0.50 & 0.47 &
    SDXL-DucHaiten3 & 0.83 & 0.50 &
    SDXL-Ultra-9 & 0.57 & 0.50 &
    SD-BasilMix & 0.33 & 0.30 &
    FLUX-TechLine & 0.57 & 0.43 \\ \hline

    SD-RV-6.0 & 0.33 & 0.27 &
    GEN-WizardReflux & 0.60 & 0.57 &
    FLUX-Chroma-Flash & 0.77 & 0.63 &
    FLUX-BNB8 & 0.73 & 0.67 &
    SD-Tiny-Rand-Safe & 0.00 & 0.00 \\ \hline

    SDXL-Illus-Early & 0.20 & 0.17 &
    SD-RV-5.1 & 0.67 & 0.57 &
    FLUX-Willow & 0.60 & 0.50 &
    SDXL-LAI & 0.63 & 0.37 &
    SD-Tiny-SD & 0.57 & 0.47 \\ \hline

    SDXL-Blacklight-LoRA & 0.37 & 0.37 &
    FLUX-AnaLora & 0.67 & 0.57 &
    SD-Hyper & 0.67 & 0.50 &
    SD-Floral & 0.47 & 0.27 &
    GEN-IF-XL1.0 & 0.57 & 0.50 \\ \hline

    FLUX-Game-LoRA & 0.60 & 0.50 &
    SDXL-M3-1 & 0.90 & 0.60 &
    SDXL-Zuki-50 & 0.50 & 0.37 &
    SDXL-MC-Skin & 0.10 & 0.10 &
    SDXL-DS-Lightning & 0.83 & 0.47 \\ \hline

    SD-DLRemix & 0.77 & 0.40 &
    SDXL-M3-3 & 0.40 & 0.20 &
    stable-cascade & 0.73 & 0.50 &
    SDXL-Nova-80 & 0.50 & 0.30 &
    SDXL-Albedo13 & 0.87 & 0.57 \\ \hline

    SD-SD3.5-Turbo & 0.77 & 0.37 &
    SD-Memento & 0.63 & 0.47 &
    SDXL-Pony-23 & 0.30 & 0.23 &
    GEN-Kandinsky2.2D & 0.70 & 0.33 &
    SDXL-M3-6 & 0.50 & 0.30 \\ \hline

    SD-EpiPhoto & 0.73 & 0.40 &
    SDXL-RV3 & 0.87 & 0.57 &
    SDXL-NoobAI-Vpred & 0.67 & 0.57 &
    SDXL-M3-0 & 0.23 & 0.23 &
    FLUX-Impress & 0.77 & 0.57 \\ \hline

    SDXL-RV5-Lightning & 0.87 & 0.63 &
    FLUX-Pixel-LoRA & 0.70 & 0.67 &
    SD-DPR2 & 0.73 & 0.47 &
    SDXL-Pony-50 & 0.37 & 0.23 &
    FLUX-Romantic-LoRA & 0.77 & 0.63 \\ \hline

    SD-DreamDiff & 0.73 & 0.17 &
    SD-DreamLike & 0.70 & 0.37 &
    FLUX-Castor3D & 0.60 & 0.60 &
    SD-SD15 & 0.37 & 0.33 &
    SD-Tiny-Rand-Safe & 0.00 & 0.00 \\ \hline

    SDXL-WAI-140 & 0.50 & 0.37 &
    SD-Tiny-Rand3 & 0.00 & 0.00 &
    SD-Canvers & 0.40 & 0.30 &
    SDXL-BigX-Tasty & 0.00 & 0.00 &
    SDXL-M3-7 & 0.87 & 0.43 \\ \hline

    FLUX-UmeSky-LoRA & 0.67 & 0.53 &
    SDXL-Illust-15 & 0.47 & 0.20 &
    SD-RV-3.0 & 0.70 & 0.60 &
    SDXL-M3-5 & 0.53 & 0.33 &
    GEN-PixArt-Sigma & 0.63 & 0.33 \\ \hline

    FLUX-Schnell & 0.63 & 0.53 &
    SDXL-LCM-LoRA & 0.77 & 0.27 &
    SD-RV-4.0 & 0.60 & 0.47 &
    SD-SDT5 & 0.07 & 0.07 &
    FLUX-ArtNouveau & 0.77 & 0.57 \\ \hline

    FLUX-Asian2 & 0.83 & 0.80 &
    FLUX-CutePuss & 0.20 & 0.13 &
    SD-Floor-LoRA & 0.37 & 0.20 &
    FLUX-Arch & 0.67 & 0.60 &
    FLUX-Krea & 0.63 & 0.53 \\ \hline

    FLUX-Anime-LoRA & 0.70 & 0.67 &
    SD-Logo-FT & 0.57 & 0.37 &
    FLUX-Logo-LoRA & 0.77 & 0.73 &
    GEN-Lumina-2.0 & 0.73 & 0.47 &
    FLUX-Panorama & 0.77 & 0.53 \\ \hline

    FLUX-Anime2 & 0.70 & 0.60 &
    SD-Ghibli & 0.30 & 0.20 &
    SD-OJ-4 & 0.63 & 0.30 &
    FLUX-Lite8B & 0.73 & 0.53 &
    FLUX-Fashion-LoRA & 0.60 & 0.47 \\ \hline

    SD-AbsReal & 0.30 & 0.27 &
    FLUX-NSFW & 0.63 & 0.60 &
    FLUX-NSFW-Master & 0.83 & 0.67 &
    SDXL-IllustMix6 & 0.17 & 0.13 &
    FLUX-AWPortrait & 0.67 & 0.33 \\ \hline

    SD-MiniSD & 0.40 & 0.10 &
    SD-Taiyi-1B & 0.07 & 0.07 &
    SD-Hassan14 & 0.40 & 0.37 &
    FLUX-Realism-LoRA & 0.80 & 0.70 &
    FLUX-NSFW-HR & 0.70 & 0.33 \\ \hline

    proteusv0.3 & 0.87 & 0.70 &
    SD-RoboDiff & 0.47 & 0.23 &
    proteusv0.2 & 0.87 & 0.57 &
    SD-DS-7 & 0.40 & 0.27 &
    SD-NSFW-Anime2 & 0.87 & 0.43 \\ \hline

    GEN-Hunyuan-DiT & 0.60 & 0.47 &
    SDXL-StyleEnh-LoRA & 0.80 & 0.47 &
    GEN-SSD-1B & 0.87 & 0.33 &
    FLUX-Chroma-HD & 0.73 & 0.67 &
    SDXL-Segmind-Vega & 0.87 & 0.47 \\ \hline

    SDXL-MWRI-NSFW & 0.63 & 0.47 &
    SDXL-PixelGen & 0.83 & 0.57 &
    SD-SD-Turbo & 0.80 & 0.07 &
    SDXL-Pornworks3 & 0.00 & 0.00 &
    snimagine-xl-3.1 & 0.47 & 0.33 \\ \hline

    SDXL-SDXL & 0.83 & 0.07 &
    SDXL-NSFW-Uncens & 0.50 & 0.23 &
    SD-SD2 & 0.70 & 0.43 &
    SD-ExprH & 0.87 & 0.53 &
    SDXL-DMD2 & 0.80 & 0.60 \\ \hline

    GEN-CogView4 & 0.57 & 0.43 &
    FLUX & 0.73 & 0.67 &
    SDXL-DS-Turbo & 0.73 & 0.57 &
    GEN-Kandinsky2.2P & 0.63 & 0.33 &
    SD-NSFW-Waifu & 0.33 & 0.30 \\ \hline

    SD-SD3.5 & 0.70 & 0.43 &
    SDXL-SDXL-Lgt & 0.57 & 0.30 &
    SD-CleanMix-NSFW & 0.53 & 0.17 &
    SD-EpicReal & 0.37 & 0.33 &
    SD-LCM-SD15-LoRA & 0.20 & 0.17 \\ \hline

    SDXL-NSFW-Gen2 & 0.90 & 0.57 &
    SDXL-Fluently4 & 0.90 & 0.47 &
    GEN-Playground2.5 & 0.50 & 0.30 &
    SDXL-Animagine-3.1 & 0.57 & 0.27 &
    SDXL-Koala-Lgt & 0.57 & 0.23 \\ \hline

    SD-DLP-2.0 & 0.67 & 0.40 &
    SD-NSFW-Gen2.1 & 0.83 & 0.50 &
    SDXL-Animagine-4.0 & 0.33 & 0.27 &
    SD-SD3 & 0.83 & 0.50 &
    SDXL-PokemonPix & 0.00 & 0.00 \\ \hline

    SD-LCM-DS7 & 0.20 & 0.10 &
    SDXL-SDXL-Turbo & 0.67 & 0.10 &
    SD-SD1.5 & 0.33 & 0.27 &
    SD-DS-8 & 0.33 & 0.23 &
    QWEN-StudioReal & 0.40 & 0.33 \\ \hline

    QWEN-Raena & 0.77 & 0.57 &
    GEN-Qwen-NSFW & 0.77 & 0.17 &
    QWEN-Lightning & 0.77 & 0.63 &
    QWEN-AWPortrait & 0.80 & 0.67 &
    QWEN-Boreal & 0.23 & 0.23 \\ \hline

    QWEN-Realism-LoRA & 0.50 & 0.40 &
    QWEN-Base & 0.87 & 0.60 &
    QWEN-HeadshotX & 0.73 & 0.53 &
    & {} & {} &
    & {} & {} \\

    \bottomrule
  \end{tabular}
\end{table*}
\small
\setlength{\tabcolsep}{3pt}

\begin{table*}[t]
    \centering
    \tiny
    \caption{Security landscape of In-the-wild T2I models using MMA.}
    \label{tab:rq2_3}
    
    \begin{tabular}{
        l|c|c||
        l|c|c||
        l|c|c||
        l|c|c||
        l|c|c
    }
    
    \toprule
    \textbf{Model} & \textbf{AASR} & \textbf{ASR} &
    \textbf{Model} & \textbf{AASR} & \textbf{ASR} &
    \textbf{Model} & \textbf{AASR} & \textbf{ASR} &
    \textbf{Model} & \textbf{AASR} & \textbf{ASR} &
    \textbf{Model} & \textbf{AASR} & \textbf{ASR} \\
    \hline
    FLUX-ArtNouveau & 0.30 & 0.37 &
    GEN-CogView4 & 0.27 & 0.30 &
    SDXL-RV5-Lightning & 0.53 & 0.60 &
    SDXL-LCM & 0.23 & 0.50 &
    FLUX-Asian2 & 0.37 & 0.37 \\ \hline

    FLUX & 0.00 & 0.00 &
    FLUX-Pixel-LoRA & 0.37 & 0.37 &
    FLUX-Ghibli-LoRA & 0.20 & 0.23 &
    FLUX-CutePuss & 0.30 & 0.43 &
    SDXL-DS-Turbo & 0.57 & 0.80 \\ \hline

    SD-DPR2 & 0.57 & 0.90 &
    SDXL-BigX-Photo & 0.00 & 0.00 &
    SD-Floor-LoRA & 0.17 & 0.27 &
    GEN-Kandinsky2.2P & 0.17 & 0.23 &
    SDXL-Pony-50 & 0.67 & 0.83 \\ \hline

    GEN-WizardReflux & 0.20 & 0.27 &
    FLUX-Arch & 0.27 & 0.37 &
    SD-NSFW-Waifu & 0.10 & 0.10 &
    FLUX-Romantic-LoRA & 0.37 & 0.43 &
    FLUX-3DRender & 0.40 & 0.57 \\ \hline

    FLUX-Krea & 0.07 & 0.13 &
    SD-SD3.5 & 0.10 & 0.17 &
    SD-DreamDiff & 0.27 & 1.00 &
    SDXL-GurilaMash & 0.83 & 1.00 &
    FLUX-Anime-LoRA & 0.23 & 0.37 \\ \hline

    SDXL-SDXL-Lgt & 0.13 & 0.30 &
    SD-DreamLike & 0.67 & 0.97 &
    FLUX-Willow & 0.17 & 0.17 &
    SD-Logo-FT & 0.07 & 0.17 &
    SDXL-AAMAnime & 0.40 & 0.73 \\ \hline

    FLUX-Castor3D & 0.53 & 0.57 &
    GEN-Tiny-Sana & 0.00 & 0.00 &
    FLUX-Logo-LoRA & 0.33 & 0.37 &
    SD-CleanMix-NSFW & 0.33 & 0.53 &
    SD-SD15 & 0.07 & 0.17 \\ \hline

    SDXL-WAI-80 & 0.57 & 0.70 &
    GEN-Lumina-2.0 & 0.37 & 0.47 &
    SD-EpicReal & 0.03 & 0.07 &
    SD-Tiny-Rand-Safe & 0.00 & 0.00 &
    SDXL-Kohaku-B5 & 0.47 & 0.93 \\ \hline

    FLUX-Panorama & 0.40 & 0.43 &
    SD-LCM-SD15-LoRA & 0.00 & 0.00 &
    SDXL-WAI-140 & 0.63 & 0.80 &
    SDXL-Cyborg-LoRA & 0.33 & 0.50 &
    FLUX-Anime2 & 0.27 & 0.40 \\ \hline

    SDXL-NSFW-Gen2 & 0.60 & 0.87 &
    SD-Tiny-Rand3 & 0.00 & 0.00 &
    FLUX-Chroma-Flash & 0.33 & 0.57 &
    SD-Ghibli & 0.03 & 0.10 &
    SDXL-Fluently4 & 0.60 & 0.90 \\ \hline

    SD-Canvers & 0.10 & 0.10 &
    GEN-ScandiInterior & 0.23 & 0.27 &
    SD-OJ-4 & 0.07 & 0.17 &
    GEN-Playground2.5 & 0.07 & 0.13 &
    SDXL-BigX-Tasty & 0.00 & 0.00 \\ \hline

    FLUX-TechLine & 0.27 & 0.30 &
    FLUX-Lite8B & 0.43 & 0.53 &
    SDXL-Animagine-3.1 & 0.13 & 0.37 &
    SDXL-M3-7 & 0.43 & 0.80 &
    SDXL-Ultra-9 & 0.63 & 0.80 \\ \hline

    FLUX-Fashion-LoRA & 0.40 & 0.43 &
    SD-OJ & 0.03 & 0.15 &
    FLUX-UmeSky-LoRA & 0.27 & 0.37 &
    GEN-EmmaWatson & 0.53 & 0.60 &
    SD-SDT5 & 0.50 & 0.67 \\ \hline

    SDXL-Koala-Lgt & 0.00 & 0.10 &
    SDXL-Illust-15 & 0.63 & 0.93 &
    FLUX-Boreal & 0.23 & 0.23 &
    SD-AbsReal & 0.03 & 0.07 &
    SD-DLP-2.0 & 0.50 & 0.97 \\ \hline

    SD-RV-3.0 & 0.87 & 1.00 &
    SD-Tiny-SD & 0.33 & 0.63 &
    FLUX-NSFW & 0.67 & 0.77 &
    SD-NSFW-Gen2.1 & 0.53 & 0.90 &
    SDXL-M3-5 & 0.63 & 0.83 \\ \hline

    SDXL-Blacklight-LoRA & 0.07 & 0.17 &
    FLUX-NSFW-Master & 0.63 & 0.77 &
    SDXL-Animagine-4.0 & 0.07 & 0.07 &
    GEN-PixArt-Sigma & 0.13 & 0.17 &
    SDXL-AIIllust & 0.47 & 0.67 \\ \hline

    SDXL-IllustMix6 & 0.13 & 0.23 &
    GEN-PixArt1024 & 0.00 & 0.03 &
    FLUX-Schnell & 0.07 & 0.07 &
    SD-Tiny-Rand-Safe & 0.00 & 0.00 &
    FLUX-AWPortrait & 0.23 & 0.43 \\ \hline

    SD-SD3 & 0.00 & 0.13 &
    SDXL-LCM-LoRA & 0.17 & 0.37 &
    GEN-Phantasma & 0.07 & 0.07 &
    SD-MiniSD & 0.00 & 0.33 &
    SDXL-PokemonPix & 0.03 & 0.03 \\ \hline

    SD-RV-4.0 & 0.87 & 1.00 &
    FLUX-Margot & 0.17 & 0.23 &
    SD-Taiyi-1B & 0.00 & 0.00 &
    SD-LCM-DS7 & 0.03 & 0.10 &
    QWEN-StudioReal & 0.07 & 0.07 \\ \hline

    SD-BasilMix & 0.00 & 0.00 &
    FLUX-SN2 & 0.67 & 0.80 &
    SDXL-SDXL-Turbo & 0.10 & 0.50 &
    QWEN-Raena & 0.13 & 0.27 &
    FLUX-SRPO & 0.27 & 0.30 \\ \hline

    FLUX-Realism-LoRA & 0.27 & 0.30 &
    SD-SD1.5 & 0.10 & 0.23 &
    GEN-Qwen-NSFW & 0.20 & 0.87 &
    SD-Anything4 & 0.00 & 0.00 &
    FLUX-NSFW-HR & 0.47 & 0.83 \\ \hline

    SD-DS-8 & 0.03 & 0.03 &
    QWEN-Lightning & 0.23 & 0.23 &
    GEN-Qwen-4bit & 0.07 & 0.10 &
    proteusv0.3 & 0.63 & 0.80 &
    SDXL-PixArt512 & 0.03 & 0.07 \\ \hline

    QWEN-AWPortrait & 0.30 & 0.40 &
    SDXL-PixelArt & 0.20 & 0.27 &
    SD-RoboDiff & 0.10 & 0.23 &
    GEN-IF-XL1.0 & 0.07 & 0.10 &
    QWEN-Boreal & 0.07 & 0.07 \\ \hline

    SDXL-Crystal & 0.33 & 0.53 &
    proteusv0.2 & 0.53 & 0.83 &
    FLUX-Game-LoRA & 0.27 & 0.33 &
    QWEN-Realism-LoRA & 0.13 & 0.17 &
    GEN-90Anime & 0.17 & 0.20 \\ \hline

    SD-DS-7 & 0.00 & 0.03 &
    SDXL-M3-1 & 0.70 & 0.93 &
    QWEN-Base & 0.20 & 0.27 &
    FLUX-AntiBlur & 0.20 & 0.27 &
    SD-NSFW-Anime2 & 0.37 & 0.80 \\ \hline

    SDXL-Zuki-50 & 0.77 & 0.87 &
    QWEN-HeadshotX & 0.07 & 0.13 &
    SDXL-StickersRed & 0.27 & 0.43 &
    SDXL-OpenDalle1.1 & 0.70 & 0.93 &
    SDXL-MC-Skin & 0.10 & 0.10 \\ \hline

    SD-RV-5.1 & 0.83 & 0.93 &
    SDXL-Disney-LoRA & 0.33 & 0.47 &
    GEN-Hunyuan-DiT & 0.20 & 0.30 &
    SDXL-DS-Lightning & 0.60 & 0.73 &
    SDXL-LAI & 0.07 & 0.07 \\ \hline

    SD-DLA-1.0 & 0.47 & 1.00 &
    SDXL-StyleEnh-LoRA & 0.45 & 0.72 &
    SD-DLP-Remix & 0.80 & 0.97 &
    SDXL-AlbedoBase & 0.60 & 0.83 &
    SDXL-RV4 & 0.67 & 0.73 \\ \hline

    SDXL-DashAnime & 0.37 & 0.80 &
    SD-DLRemix & 0.53 & 1.00 &
    FLUX-RetroAnime & 0.20 & 0.33 &
    SD-SD3-Tiny & 0.00 & 0.00 &
    GEN-SSD-1B & 0.27 & 0.57 \\ \hline

    SDXL-M3-3 & 0.03 & 0.03 &
    SD-Floral & 0.20 & 0.30 &
    FLUX-Shuttle3 & 0.13 & 0.27 &
    FLUX-Chroma-HD & 0.40 & 0.53 &
    GEN-StableCascade & 0.07 & 0.07 \\ \hline

    SD-Hyper & 0.37 & 0.40 &
    GEN-Kandinsky2.1 & 0.13 & 0.13 &
    SDXL-Segmind-Vega & 0.40 & 0.73 &
    SDXL-Nova-80 & 0.67 & 0.73 &
    FLUX-Megan & 0.47 & 0.53 \\ \hline

    SD-Inkpunk & 0.00 & 0.00 &
    SDXL-AnimeDetail-LoRA & 0.45 & 0.55 &
    SDXL-Albedo13 & 0.60 & 0.90 &
    SD-DreamShaper & 0.03 & 0.03 &
    SDXL-Juggernaut-XI & 0.73 & 0.93 \\ \hline

    SDXL-MWRI-NSFW & 0.80 & 1.00 &
    SD-SD3.5-Turbo & 0.20 & 0.27 &
    FLUX-Ghibsky & 0.27 & 0.30 &
    GEN-Tiny-LCM & 0.00 & 0.00 &
    SDXL-PixelGen & 0.50 & 0.80 \\ \hline

    SD-Memento & 0.77 & 1.00 &
    FLUX-TestLLM & 0.30 & 0.33 &
    SD-DLP-1.0 & 0.77 & 1.00 &
    GEN-PixArt-Sigma & 0.03 & 0.17 &
    SDXL-Pony-23 & 0.77 & 0.83 \\ \hline

    FLUX-Point & 0.27 & 0.30 &
    SDXL-Animagine-2.0 & 0.33 & 0.67 &
    SD-SD-Turbo & 0.00 & 0.47 &
    GEN-Kandinsky2.2D & 0.10 & 0.13 &
    SDXL-3DRender & 0.13 & 0.67 \\ \hline

    GEN-DronePhoto & 0.33 & 0.63 &
    SDXL-Pornworks3 & 0.00 & 0.00 &
    SDXL-M3-6 & 0.63 & 0.90 &
    FLUX-Chroma-Base & 0.40 & 0.43 &
    SDXL-DucHaiten3 & 0.70 & 0.97 \\ \hline

    SDXL-SDXL & 0.03 & 0.40 &
    SD-EpiPhoto & 0.60 & 1.00 &
    SDXL-JankuV5 & 0.60 & 0.83 &
    SD-Counterfeit25 & 0.20 & 0.43 &
    SDXL-NSFW-Uncens & 0.57 & 0.83 \\ \hline

    SDXL-RV3 & 0.53 & 0.70 &
    FLUX-Turbo-Alpha & 0.23 & 0.30 &
    GEN-FilmPortrait & 0.13 & 0.13 &
    SD-SD1.4 & 0.13 & 0.13 &
    SDXL-NoobAI-1.1 & 0.00 & 0.00 \\ \hline

    FLUX-BNB8 & 0.33 & 0.43 &
    GEN-80Fantasy & 0.50 & 0.53 &
    SD-SD2 & 0.00 & 0.03 &
    SDXL-NoobAI-Vpred & 0.60 & 0.67 &
    SDXL-Illus-Early & 0.23 & 0.37 \\ \hline

    SD-ExprH & 0.23 & 0.27 &
    SDXL-M3-0 & 0.73 & 0.77 &
    SD-RV-6.0 & 0.03 & 0.03 &
    SDXL-DMD2 & 0.37 & 0.43 &
    FLUX-Impress & 0.40 & 0.40 \\ \hline

    GEN-Storyboard & 0.03 & 0.10 &
    & {} & {} &
    & {} & {} &
    & {} & {} &
    & {} & {} \\

    \bottomrule
  \end{tabular}
\end{table*}
\small
\setlength{\tabcolsep}{3pt}

\begin{table*}[t]
    \centering
    \tiny
    \caption{Security landscape of In-the-wild T2I models using 4chan jailbreak dataset.}
    \label{tab:rq2_2}
    
    \begin{tabular}{
        l|c|c||
        l|c|c||
        l|c|c||
        l|c|c||
        l|c|c
    }
    
    \toprule
    \textbf{Model} & \textbf{AASR} & \textbf{ASR} &
    \textbf{Model} & \textbf{AASR} & \textbf{ASR} &
    \textbf{Model} & \textbf{AASR} & \textbf{ASR} &
    \textbf{Model} & \textbf{AASR} & \textbf{ASR} &
    \textbf{Model} & \textbf{AASR} & \textbf{ASR} \\
    \hline
    SD-SD-Turbo & 0.00 & 0.13 &
    SDXL-SDXL & 0.00 & 0.27 &
    SD-SD3.5 & 0.23 & 0.30 &
    SD-SD3 & 0.13 & 0.20 &
    SD-SDT5 & 0.13 & 0.17 \\ \hline

    SD-SD1.5 & 0.03 & 0.13 &
    SD-DS-7 & 0.10 & 0.20 &
    SD-NSFW-Anime2 & 0.27 & 0.43 &
    SDXL-OpenDalle1.1 & 0.17 & 0.33 &
    GEN-Hunyuan-DiT & 0.27 & 0.33 \\ \hline

    SDXL-StyleEnh-LoRA & 0.12 & 0.17 &
    SDXL-DashAnime & 0.03 & 0.20 &
    FLUX-Chroma-HD & 0.10 & 0.13 &
    SDXL-Segmind-Vega & 0.10 & 0.23 &
    SDXL-AnimeDetail-LoRA & 0.10 & 0.13 \\ \hline

    SDXL-MWRI-NSFW & 0.63 & 0.77 &
    SDXL-PixelGen & 0.23 & 0.37 &
    SDXL-Pornworks3 & 0.00 & 0.00 &
    SDXL-NSFW-Uncens & 0.37 & 0.47 &
    SD-SD1.4 & 0.03 & 0.10 \\ \hline

    SD-SD2 & 0.00 & 0.03 &
    SD-ExprH & 0.17 & 0.17 &
    SDXL-DMD2 & 0.20 & 0.20 &
    GEN-CogView4 & 0.17 & 0.23 &
    SDXL-DS-Turbo & 0.20 & 0.33 \\ \hline

    GEN-Kandinsky2.2P & 0.10 & 0.13 &
    SD-NSFW-Waifu & 0.00 & 0.00 &
    SDXL-SDXL-Lgt & 0.10 & 0.10 &
    SD-CleanMix-NSFW & 0.17 & 0.33 &
    SD-EpicReal & 0.10 & 0.13 \\ \hline

    SD-LCM-SD15-LoRA & 0.07 & 0.10 &
    SDXL-NSFW-Gen2 & 0.23 & 0.27 &
    SDXL-Fluently4 & 0.23 & 0.33 &
    SDXL-Animagine-3.1 & 0.10 & 0.20 &
    SD-OJ & 0.05 & 0.15 \\ \hline

    SDXL-Koala-Lgt & 0.03 & 0.13 &
    SD-DLP-2.0 & 0.27 & 0.47 &
    SD-NSFW-Gen2.1 & 0.27 & 0.40 &
    SDXL-Animagine-4.0 & 0.03 & 0.03 &
    SDXL-PokemonPix & 0.00 & 0.00 \\ \hline

    SD-LCM-DS7 & 0.07 & 0.07 &
    SD-DS-8 & 0.10 & 0.10 &
    FLUX-NSFW & 0.13 & 0.20 &
    FLUX-NSFW-Master & 0.40 & 0.47 &
    SDXL-IllustMix6 & 0.10 & 0.17 \\ \hline

    FLUX-AWPortrait & 0.03 & 0.07 &
    SD-MiniSD & 0.03 & 0.13 &
    SD-Taiyi-1B & 0.00 & 0.00 &
    SD-Hassan14 & 0.07 & 0.13 &
    FLUX-SN2 & 0.10 & 0.17 \\ \hline

    FLUX-Realism-LoRA & 0.07 & 0.07 &
    FLUX-NSFW-HR & 0.50 & 0.93 &
    SDXL-Proteus0.3 & 0.33 & 0.40 &
    SD-RoboDiff & 0.03 & 0.07 &
    SDXL-Proteus0.2 & 0.27 & 0.37 \\ \hline

    FLUX-ArtNouveau & 0.00 & 0.03 &
    FLUX-Asian2 & 0.17 & 0.17 &
    FLUX-CutePuss & 0.10 & 0.17 &
    SD-Floor-LoRA & 0.03 & 0.10 &
    FLUX-Arch & 0.10 & 0.13 \\ \hline

    FLUX-Krea & 0.03 & 0.03 &
    FLUX-Anime-LoRA & 0.10 & 0.17 &
    SD-Logo-FT & 0.03 & 0.07 &
    FLUX-Logo-LoRA & 0.10 & 0.10 &
    FLUX-Panorama & 0.07 & 0.07 \\ \hline

    FLUX-Anime2 & 0.07 & 0.13 &
    SD-Ghibli & 0.07 & 0.07 &
    SD-OJ-4 & 0.03 & 0.10 &
    FLUX-Lite8B & 0.20 & 0.20 &
    FLUX-Fashion-LoRA & 0.17 & 0.17 \\ \hline

    SD-AbsReal & 0.13 & 0.17 &
    GEN-IF-XL1.0 & 0.07 & 0.10 &
    FLUX-Game-LoRA & 0.10 & 0.13 &
    SDXL-M3-1 & 0.20 & 0.37 &
    SDXL-Zuki-50 & 0.43 & 0.63 \\ \hline

    SDXL-MC-Skin & 0.07 & 0.07 &
    SDXL-DS-Lightning & 0.17 & 0.27 &
    SD-DLP-Remix & 0.33 & 0.37 &
    SD-DLRemix & 0.17 & 0.37 &
    SDXL-M3-3 & 0.07 & 0.10 \\ \hline

    GEN-StableCascade & 0.17 & 0.20 &
    SDXL-Nova-80 & 0.60 & 0.70 &
    SDXL-Albedo13 & 0.30 & 0.40 &
    SD-SD3.5-Turbo & 0.13 & 0.20 &
    SD-Memento & 0.33 & 0.50 \\ \hline

    SDXL-Pony-23 & 0.43 & 0.50 &
    GEN-Kandinsky2.2D & 0.03 & 0.07 &
    SDXL-M3-6 & 0.33 & 0.60 &
    SD-EpiPhoto & 0.27 & 0.40 &
    SDXL-RV3 & 0.23 & 0.27 \\ \hline

    SDXL-NoobAI-1.1 & 0.17 & 0.37 &
    SDXL-NoobAI-Vpred & 0.40 & 0.50 &
    SDXL-M3-0 & 0.23 & 0.23 &
    FLUX-Impress & 0.20 & 0.23 &
    SDXL-RV5-Lightning & 0.27 & 0.30 \\ \hline

    FLUX-Pixel-LoRA & 0.13 & 0.13 &
    SD-DPR2 & 0.27 & 0.40 &
    SDXL-Pony-50 & 0.40 & 0.60 &
    FLUX-Romantic-LoRA & 0.13 & 0.17 &
    SD-DreamDiff & 0.10 & 0.27 \\ \hline

    SD-DreamLike & 0.27 & 0.37 &
    FLUX-Castor3D & 0.13 & 0.13 &
    SD-SD15 & 0.10 & 0.10 &
    SD-Tiny-Rand-Safe & 0.00 & 0.00 &
    SDXL-WAI-140 & 0.37 & 0.50 \\ \hline

    SD-Tiny-Rand3 & 0.00 & 0.00 &
    SD-Canvers & 0.13 & 0.17 &
    SDXL-BigX-Tasty & 0.00 & 0.00 &
    SDXL-M3-7 & 0.20 & 0.33 &
    FLUX-UmeSky-LoRA & 0.03 & 0.07 \\ \hline

    SDXL-Illust-15 & 0.43 & 0.77 &
    SD-RV-3.0 & 0.13 & 0.30 &
    SDXL-M3-5 & 0.27 & 0.47 &
    GEN-PixArt-Sigma & 0.13 & 0.33 &
    FLUX-Schnell & 0.13 & 0.13 \\ \hline

    SDXL-LCM-LoRA & 0.07 & 0.13 &
    SD-RV-4.0 & 0.33 & 0.37 &
    QWEN-StudioReal & 0.10 & 0.10 &
    QWEN-Raena & 0.00 & 0.00 &
    GEN-Qwen-NSFW & 0.17 & 0.53 \\ \hline

    QWEN-Lightning & 0.17 & 0.17 &
    QWEN-AWPortrait & 0.07 & 0.10 &
    QWEN-Boreal & 0.03 & 0.03 &
    QWEN-Realism-LoRA & 0.00 & 0.00 &
    QWEN-Base & 0.07 & 0.13 \\ \hline

    QWEN-HeadshotX & 0.07 & 0.13 &
    SD-SD3-Tiny & 0.00 & 0.00 &
    FLUX-AntiBlur & 0.10 & 0.10 &
    FLUX-SRPO & 0.17 & 0.17 &
    FLUX-TestLLM & 0.13 & 0.13 \\ \hline

    GEN-Phantasma & 0.00 & 0.00 &
    FLUX-Boreal & 0.03 & 0.03 &
    SD-Counterfeit25 & 0.03 & 0.10 &
    SDXL-Crystal & 0.20 & 0.23 &
    FLUX-Chroma-Base & 0.10 & 0.13 \\ \hline

    SD-DLA-1.0 & 0.20 & 0.37 &
    SDXL-Juggernaut-XI & 0.33 & 0.40 &
    SDXL-BigX-Photo & 0.00 & 0.00 &
    SDXL-Disney-LoRA & 0.03 & 0.07 &
    SDXL-GurilaMash & 0.53 & 0.60 \\ \hline

    FLUX-Megan & 0.10 & 0.10 &
    FLUX-Shuttle3 & 0.20 & 0.20 &
    FLUX-Ghibsky & 0.03 & 0.03 &
    SDXL-AIIllust & 0.13 & 0.17 &
    SDXL-Animagine-2.0 & 0.17 & 0.20 \\ \hline

    GEN-ScandiInterior & 0.13 & 0.13 &
    SDXL-StickersRed & 0.10 & 0.20 &
    GEN-FilmPortrait & 0.00 & 0.00 &
    SD-DLP-1.0 & 0.20 & 0.40 &
    SD-Inkpunk & 0.00 & 0.00 \\ \hline

    SDXL-AlbedoBase & 0.20 & 0.33 &
    GEN-Qwen-4bit & 0.00 & 0.00 &
    SDXL-3DRender & 0.07 & 0.13 &
    SDXL-LCM & 0.17 & 0.17 &
    SDXL-JankuV5 & 0.47 & 0.53 \\ \hline

    SDXL-RV5 & 0.23 & 0.30 &
    FLUX-Turbo-Alpha & 0.17 & 0.17 &
    FLUX-Ghibli-LoRA & 0.27 & 0.27 &
    GEN-DronePhoto & 0.07 & 0.17 &
    SDXL-Kohaku-B5 & 0.20 & 0.23 \\ \hline

    GEN-Kandinsky2.1 & 0.27 & 0.27 &
    SD-DreamShaper & 0.03 & 0.03 &
    SDXL-RV4 & 0.17 & 0.20 &
    SDXL-DucHaiten3 & 0.40 & 0.53 &
    SDXL-Ultra-9 & 0.33 & 0.37 \\ \hline

    SD-BasilMix & 0.00 & 0.03 &
    SD-RV-6.0 & 0.10 & 0.10 &
    FLUX-Chroma-Flash & 0.17 & 0.20 &
    FLUX-BNB8 & 0.13 & 0.13 &
    SD-Tiny-Rand-Safe & 0.00 & 0.00 \\ \hline

    SDXL-Illus-Early & 0.20 & 0.23 &
    SD-RV-5.1 & 0.37 & 0.43 &
    SDXL-LAI & 0.10 & 0.17 &
    SD-Tiny-SD & 0.23 & 0.27 &
    SDXL-Blacklight-LoRA & 0.07 & 0.07 \\ \hline

    SD-Hyper & 0.27 & 0.27 &
    SD-Floral & 0.03 & 0.17 &
    & {} & {} &
    & {} & {} &
    & {} & {} \\
    \bottomrule
  \end{tabular}
\end{table*}

\end{document}